\documentclass[fleqn,10pt]{wlscirep}
\usepackage[utf8]{inputenc}
\usepackage[T1]{fontenc}
\usepackage{bm}
\usepackage{float}
\usepackage{braket}
\usepackage{multirow}
\usepackage{tikz}
\usepackage{quantikz}
\usepackage{placeins}
\usetikzlibrary{positioning,calc,arrows.meta,shapes.geometric}
\usetikzlibrary{positioning}
\usepackage[normalem]{ulem}
\usepackage{xcolor}

\title{An Analytically Trained Variational Quantum Circuit using the Output Probability Distribution of Quantum Phase Estimation on NISQ Hardware}

\author[1, 2]{\small Mousumi Kundu}
\author[1]{\small Ashish Kumar Patra}
\author[1]{\small Anurag K. S. V.}
\author[1, 3]{\small Ruchika Bhat}
\author[1]{\small Sai Shankar P.}
\author[4]{\small Alok Shukla}
\author[, 1]{\small Jaiganesh G.\thanks{(Corresponding Author) email: jaiganesh@qclairvoyance.in, drjaiganesh15@gmail.com}}

\affil[1]{\small Qclairvoyance Quantum Labs, Secunderabad, TG 500094, India.}
\affil[2]{\small Department of CSE(AI \& ML), School of Engineering, Dayananda Sagar University, Bengaluru, KA 562112, India.}
\affil[3]{\small The University of Arizona, Tucson, AZ 85721, USA.}
\affil[4]{\small School of Arts and Sciences, Ahmedabad University, Ahmedabad, GJ 380009, India.}

\begin{abstract}

Quantum Phase Estimation (QPE) is the standard algorithm for molecular ground-state energy
estimation and often serves as a subroutine in larger quantum algorithms, but its deep
circuits make it impractical on current noisy hardware. A previous variational approach
replaces the QPE subroutine with a shallow Variational Quantum Circuit (VQC) trained to
reproduce the QPE measurement distribution, but generates the training target by classically
simulating the full QPE circuit, which is prohibitive for molecules. We reformulate
this idea, computing the training target analytically from the Dirichlet kernel, using only
a reference energy, the number of ancilla qubits and the evolution time, so the
controlled-unitary and inverse-QFT operations of QPE are never simulated. Using H$_2$, we
select the VQC stepwise by comparing its IBM Quantum hardware output with the target using
four metrics. We first choose the entangler (linear or full) and number of layers of an
$R_Y$-$R_Z$-$CZ$ ansatz, with and without dynamical decoupling (DD). As DD degraded
performance with depth, we then choose, without DD, the ansatz ($R_Y$-$R_Z$-$CZ$ or
$R_Y$-$CZ$) and the training simulator (ideal or noisy). The configuration fixed on H$_2$, a
single-layer linear-entangler $R_Y$-$CZ$ ansatz trained on the ideal simulator, is validated
on LiH, H$_2$O, BeH$_2$, NH$_3$, CH$_4$, O$_2$ and N$_2$, recovering the dominant QPE
bitstring on hardware with errors within chemical accuracy ($1.59\times10^{-3}$ Ha) for all
eight molecules at the chosen evolution times. We also analyze the hardware side lobes of
both ans\"atze, and a noise analysis of the selected ansatz at 8 and 64 layers examines how DD
performs with depth. As training requires a known reference energy, the surrogate is not
meant for standalone energy estimation, but can replace the QPE subroutine where a larger
algorithm uses only its measurement outcome.
\end{abstract}

\begin{document}

\flushbottom
\maketitle

\thispagestyle{empty}

\noindent \textbf{Keywords:}
Quantum Phase Estimation · Variational Quantum Circuits · Quantum Chemistry · Dirichlet Kernel · Molecular Energy Estimation

\newpage

\section{Introduction}
\label{sec:introduction}

Quantum computing has emerged as a transformative paradigm for addressing
computational problems that remain intractable for classical machines,
leveraging the principles of superposition, entanglement, and quantum
interference~\cite{nielsen2000quantum, preskill2018quantum,
steane1998quantum}. Among the foundational subroutines underpinning many
quantum algorithms, Quantum Phase Estimation (QPE) occupies a central
role, providing the mechanism by which the eigenphase of a unitary
operator is extracted with high precision~\cite{kitaev1995quantum,
cleve1998quantum, nielsen2000quantum}. Originally introduced by
Kitaev~\cite{kitaev1995quantum} and later refined into the circuit-based
formulation by Cleve~\textit{et al.}~\cite{cleve1998quantum}, QPE has
since become an indispensable primitive in quantum algorithm design. It
serves as a critical component in landmark algorithms such as Shor's
factoring algorithm~\cite{shor1994algorithms} and the HHL algorithm for
linear systems of equations~\cite{harrow2009quantum}, and it is the
standard quantum algorithm for molecular ground-state energy
estimation~\cite{aspuru2005simulated, babbush2018encoding,
reiher2017elucidating}. Molecular energy estimation by QPE in turn
appears as a subroutine within larger computational workflows, such as
the study of reaction mechanisms and thermochemistry, in which QPE
energies of intermediates and transition states are combined into
reaction energies and barriers~\cite{reiher2017elucidating}, and the
calculation of energy derivatives for molecular properties and geometry
optimization~\cite{kassal2009quantum, obrien2019calculating}.

Despite its theoretical power, the standard QPE circuit incurs a deep
circuit requirement that renders it acutely sensitive to gate errors and
decoherence, posing a fundamental obstacle in the Noisy
Intermediate-Scale Quantum (NISQ)
era~\cite{preskill2018quantum, leymann2020bitter, bharti2022noisy}. The
accumulation of noise over many sequential gate operations degrades the
fidelity of measurement outcomes, in some cases to the point where the
dominant bitstring is irrecoverably
corrupted~\cite{temme2017error, endo2021hybrid}. Various strategies have
been explored to address this challenge, including error mitigation
techniques~\cite{temme2017error, li2017efficient, giurgica2020digital},
compilation optimization~\cite{lao2022timing}, and hybrid
quantum-classical algorithms~\cite{cerezo2021variational,
mcclean2016theory} designed to minimize circuit depth while preserving
computational utility.

Among hybrid approaches, Variational Quantum Algorithms (VQAs) have
attracted considerable attention as a near-term alternative to deep
quantum circuits~\cite{cerezo2021variational, bharti2022noisy}. The
Variational Quantum Eigensolver (VQE), introduced by Peruzzo
\textit{et al.}~\cite{peruzzo2014variational} and subsequently developed
extensively~\cite{kandala2017hardware, tilly2022vqe, mcclean2016theory},
estimates molecular ground-state energies using shallow parameterized
circuits optimized by classical routines. While VQE circumvents the depth
requirements of QPE, it returns an energy expectation value rather than
the QPE measurement distribution, and therefore cannot directly take the
place of a QPE block whose output is consumed by a larger quantum
algorithm.

A conceptually distinct approach to circumventing the depth limitation of QPE while
preserving its output was proposed by Liu \textit{et al.}~\cite{liu2024learnqpe}, who treat
QPE explicitly as a subroutine and show that a Variational Quantum Circuit (VQC) can be
trained to reproduce its measurement probability distribution, casting phase estimation as
a supervised learning problem and reducing the circuit depth to $\mathcal{O}(pn)$ for a VQC
with $p$ layers and $n$ qubits under a linear entangler topology. The trained VQC is then
proposed as a hardware-amenable replacement for the QPE block within a larger quantum
circuit. In that approach, however, the training target is the measurement histogram of the
full QPE circuit executed on an ideal classical simulator, which, for molecular energy
estimation, would require simulating deep controlled time-evolution operators across the
system and ancilla registers at a memory cost that grows exponentially with the total number
of qubits.

The present work is therefore not an extension of that framework but a
reformulation of the same as we retain its premise, a shallow VQC trained to reproduce the QPE
output distribution, and redesign how the training target is obtained. For an ideal QPE with
the system register prepared in the ground state, the post-inverse-QFT distribution of the
ancilla register is given in closed form by the squared Dirichlet
kernel~\cite{cleve1998quantum, nielsen2000quantum}. Given a reference ground-state energy,
the number of ancilla qubits $n_a$ and the evolution time $\tau$, the target
$P_\text{gt}(k;\phi)$ is evaluated exactly by classical arithmetic, so that neither the
controlled-unitary operations nor the inverse quantum Fourier transform of QPE is ever
constructed or simulated, and the cost of building the target shifts from simulating a QPE
circuit to obtaining the reference energy. The VQC is trained against this analytical target
and executed on IBM Quantum superconducting hardware.

To identify a suitable surrogate circuit, we carry out a stepwise
selection on the hydrogen molecule (H$_2$), in which each choice is made
by comparing the hardware-executed VQC distribution with the Dirichlet
target using four complementary metrics: the Hellinger
distance~\cite{hellinger1909neue}, fidelity error, total variation
distance (TVD), and Jensen--Shannon divergence
(JSD)~\cite{endres2003new}. We first compare linear and full entangler
topologies for an $R_Y$-$R_Z$-$CZ$ ansatz and then vary the number of
layers of the selected topology from $p = 1$ to $p = 5$; both steps are
performed with and without XpXm Dynamical Decoupling
(DD)~\cite{viola1998dynamical, viola1999dynamical,
zanardi1999symmetrizing}. Since DD degraded the performance of the
linear-entangler circuits in these experiments, the subsequent steps are carried out without DD: with the
selected entangler and depth, we compare the $R_Y$-$R_Z$-$CZ$ ansatz with
a reduced $R_Y$-$CZ$ ansatz, and finally compare parameters trained on an
ideal simulator with parameters trained on a noisy simulator. The
resulting configuration, a single-layer linear-entangler $R_Y$-$CZ$
ansatz with ideal-simulator-trained parameters, is then validated on
LiH, H$_2$O, BeH$_2$, NH$_3$, CH$_4$, O$_2$ and N$_2$. For every
molecule, the ideal-simulator and hardware executions recover the
dominant bitstring of the Dirichlet target, and the corresponding energy
lies within chemical accuracy of the full configuration interaction
(FCI) reference for the chosen evolution time.

Beyond the dominant bitstring, we analyse the side-lobe structure of the
hardware-executed distributions of both ans\"atze relative to the
Dirichlet target. In addition, we carry out a noise analysis of the
selected $R_Y$-$CZ$ ansatz at $p \in \{8, 64\}$, with and without DD,
to characterize how the effectiveness of DD depends on circuit depth on
current superconducting hardware. Because training requires the
reference energy in advance, the surrogate is not intended for
standalone energy estimation; its intended use is to replace the QPE
subroutine where only its measured output is used by a larger
algorithm.

The remainder of this paper is organized as follows.
Section~\ref{sec:methodology} describes the methodology, including the
molecular Hamiltonian construction, the analytical training pipeline,
the ansatz architectures, and the hardware execution setup.
Section~\ref{sec:results} presents and discusses the results of the
stepwise selection on H$_2$ and the validation on the remaining
molecules. Section~\ref{sec:conclusion} summarizes our findings and
outlines directions for future research. The Supplementary Material
contains three sections: Section~\ref{sec:theory} provides the theoretical
background, covering the QPE formalism and its depth scaling, the
Dirichlet-kernel target, variational quantum circuits, the cost function,
the distributional distance metrics, the molecular Hamiltonian and qubit
mapping, the molecular systems and simulation parameters, and dynamical
decoupling; Section~\ref{sec:sidelobes} presents the side-lobe structure analysis; and
Section~\ref{sec:noise_analysis} presents the noise analysis of circuit depth and dynamical
decoupling.

\section{Methodology}
\label{sec:methodology}

\subsection{Molecular Systems and Hamiltonian Construction}
\label{subsec:mol_system}

The hydrogen molecule (H$_2$) at an internuclear separation of
$R = 0.74$~\AA, close to its experimental equilibrium bond length, is
used as the benchmark system for selecting the VQC configuration. The
selected configuration is then validated on seven further molecules,
LiH, H$_2$O, BeH$_2$, NH$_3$, CH$_4$, O$_2$ and N$_2$, whose geometries
are the CCSD/STO-3G-optimized structures of Ref.~\cite{anurag2026}. The
Cartesian coordinates of all molecules are listed in Table~A2 of the
Supplementary Material. All molecules are neutral and are treated as
closed-shell singlets, except O$_2$, which is treated in its triplet
ground state.

For each molecule, the electronic Hamiltonian is constructed in the
STO-3G minimal basis set~\cite{hehre1969self} using the \texttt{PySCF}
quantum chemistry package~\cite{sun2018pyscf}, which computes the one-
and two-electron integrals $h_{pq}$ and $h_{pqrs}$ from a Hartree--Fock
reference (restricted for the closed-shell molecules and
restricted open-shell for O$_2$). The reference ground-state energy
$E_\text{FCI}$, which includes the nuclear repulsion energy
$E_\text{nuc}$, is obtained with the full configuration interaction
(FCI) solver implemented in \texttt{PySCF}~\cite{sun2018pyscf}.

The fermionic Hamiltonian is mapped to a qubit operator via the
Jordan--Wigner (JW) transformation~\cite{jordan1928paulische} using the
\texttt{Qiskit Nature} package~\cite{qiskit2023}, without further qubit
reduction. The mapped electronic Hamiltonian can be written as
\begin{equation}
    \hat{H}_\text{JW} = c_0\,\hat{I} + \hat{H}',
    \label{eq:jw_ham}
\end{equation}
where $c_0$ is the coefficient of the identity term and $\hat{H}'$
collects all non-trivial Pauli strings. Since the QPE circuit is never
constructed in the present framework, the qubit Hamiltonian enters the
training pipeline only through $c_0$, which is used to define the
shifted energy in Section~\ref{subsec:analytical_target}. 

\subsection{Analytical Training Target via the Dirichlet Kernel}
\label{subsec:analytical_target}

The central methodological device of our framework is that the QPE
simulation step used to generate training data is replaced entirely by
a closed-form evaluation of the Dirichlet kernel, derived in
Section~\ref{subsec:dirichlet} of the Supplementary Material. Rather than executing any
quantum circuit to obtain the target distribution, we compute it
directly from the reference energy, the number of ancilla qubits $n_a$,
and the evolution time $\tau$. The QPE unitary in the molecular
simulation context is the time-evolution operator
$U = e^{-i\hat{H}\tau}$, whose eigenphase encodes the energy. Since the
constant terms $c_0$ and $E_\text{nuc}$ contribute only a global phase,
the phase used for training is computed from the shifted energy
\begin{equation}
    E' = E_\text{FCI} - c_0 - E_\text{nuc},
    \label{eq:e_shift}
\end{equation}
which is the ground-state eigenvalue of $\hat{H}'$. The phase and its
rescaled form are then
\begin{equation}
    \theta = \frac{|E'|\,\tau}{2\pi}, \qquad \phi = 2^{n_a}\theta,
    \label{eq:phi_def}
\end{equation}
where the absolute value is taken because $E'$ is negative for all
molecules considered here. The target distribution is evaluated for all
$k \in \{0, 1, \ldots, 2^{n_a} - 1\}$ via the squared Dirichlet kernel,
\begin{equation}
    P_\text{gt}(k;\phi) = \frac{1}{N^2}
    \left(\frac{\sin(\pi N \delta_k)}{\sin(\pi \delta_k)}\right)^2,
    \qquad \delta_k = \theta - \frac{k}{N},
    \quad N = 2^{n_a},
    \label{eq:dirichlet_explicit}
\end{equation}
which is equivalent to Eq.~(A8) of the Supplementary Material, since
$N\delta_k = \phi - k$. The limiting value $P_\text{gt}(k;\phi) = 1$ is
assigned when $|\sin(\pi \delta_k)| < 10^{-10}$ to handle the degenerate
case $\delta_k \approx 0$, and the distribution is subsequently
normalized to ensure $\sum_k P_\text{gt}(k;\phi) = 1$.

In all experiments, we use $n_a = 11$ ancilla qubits. For H$_2$, the
evolution time is $\tau = 1.0$~a.u., which places the dominant bitstring
at $k^* = \lfloor \phi \rceil = \texttt{00101010011}$ with $98.9\%$ of
the total probability mass. For the remaining molecules, $\tau$ is
chosen individually; the values are listed in Table~A1 of the
Supplementary Material. All values satisfy $|E'|\tau < 2\pi$, so that
the phase does not wrap around, and they were selected empirically
using the FCI reference energy such that $\phi$ lies close to an
integer. The full evaluation of $P_\text{gt}(k;\phi)$ over all
$2^{11} = 2048$ bitstrings is performed entirely classically in
negligible computation time, without simulating the QPE circuit. This is
the feature that distinguishes our training pipeline from a
QPE-simulation approach: once the reference energy is available, the
cost of constructing the target does not grow with the size of the
molecule, whereas classically simulating the QPE circuit requires memory
exponential in the total number of system and ancilla qubits.

The energy recovered from the dominant bitstring $k^*$ is
\begin{equation}
    E_\text{total} = -\frac{2\pi\, \theta^*}{\tau} + c_0 + E_\text{nuc},
    \label{eq:energy_recovery_correct}
\end{equation}
where $\theta^* = k^*/N$ is the estimated phase, and the sign and
constant shifts restore the total molecular energy. The absolute energy
error is
\begin{equation}
    |\Delta E| = |E_\text{FCI} - E_\text{total}|.
    \label{eq:energy_error}
\end{equation}
Since $E_\text{total}$ depends only on $k^*$, this error equals the
discretization error $2\pi|\phi - k^*|/(2^{n_a}\tau)$ of the
$n_a$-qubit phase register whenever the measured dominant bitstring
coincides with that of the Dirichlet target.

\begin{figure*}[htbp]
\centering

\begin{tikzpicture}[
    >=Latex,
    every node/.style={font=\normalsize},
    arr/.style={-{Latex[length=3mm]}, thick},
    lbl/.style={font=\small, align=center},
    stepbox/.style={draw=black!40, rounded corners=6pt, thick, inner sep=9pt,
                     text width=4.8cm, align=left, font=\small}
]

\coordinate (leftX) at (-8.6,0);

% ================================================================
%  STEP 1: H2 MOLECULE  (text left, diagram right)
% ================================================================
\node (molecule) at (0,0) {
  \begin{tikzpicture}
    \draw[line width=4.5pt, gray!55, line cap=round] (0.08,0) -- (0.87,0);
    \shade[ball color=gray!45] (0,0) circle (0.34);
    \shade[ball color=gray!45] (0.95,0) circle (0.34);
    \node[font=\small] at (0,0) {H};
    \node[font=\small] at (0.95,0) {H};
  \end{tikzpicture}
};
\coordinate (moltextPos) at (leftX |- molecule.north);
\node[stepbox, fill=blue!7, draw=blue!35, anchor=north west] (moltext) at (moltextPos)
  {\textbf{Step 1: Quantum chemistry}\\[2pt]
  Build the molecular Hamiltonian (STO-3G, JW mapping) and compute the reference energy $E_{\mathrm{FCI}}$ via Full Configuration Interaction.\\[6pt]
  $\hat{H} = \displaystyle\sum_{p,q} h_{pq}\, \hat{a}_p^\dagger \hat{a}_q$\\[3pt]
  $+\ \dfrac{1}{2}\displaystyle\sum_{p,q,r,s} h_{pqrs}\, \hat{a}_p^\dagger \hat{a}_q^\dagger \hat{a}_r \hat{a}_s + E_{\mathrm{nuc}}$};

% robust vertical anchor: whichever is lower, the icon or its text box
\path let \p1=(molecule.south), \p2=(moltext.south) in
  \pgfextra{\pgfmathparse{min(\y1,\y2)}\xdef\stepOneBottomY{\pgfmathresult}};
\coordinate (yRefStep1) at (0,\stepOneBottomY pt);
\coordinate (step1Bottom) at (molecule |- yRefStep1);

% ================================================================
%  STEP 2: ANALYTICAL DIRICHLET TARGET DISTRIBUTION (text left, plot right)
% ================================================================
\node (dist) at ($(step1Bottom)+(0,-1.9)$) {
  \begin{tikzpicture}
    \path[use as bounding box] (-2.15,-0.25) rectangle (2.15,2.1);
    \draw[->] (-1.7,0) -- (1.9,0) node[right] {\small $k$};
    \draw[->] (0,-0.1) -- (0,1.65) node[above] {\small $P(k;\phi)$};
    \draw[thick, blue!70!black, smooth] plot coordinates {
      (-1.65,0.02) (-1.45,0.06) (-1.28,0.01) (-1.1,0.10) (-0.98,0.02)
      (-0.85,0.20) (-0.72,0.02) (-0.58,0.40) (-0.45,0.01)
      (-0.27,0.85) (-0.13,1.30) (0,1.6) (0.13,1.30) (0.27,0.85)
      (0.45,0.01) (0.58,0.40) (0.72,0.02) (0.85,0.20) (0.98,0.02)
      (1.1,0.10) (1.28,0.01) (1.45,0.06) (1.65,0.02)
    };
  \end{tikzpicture}
};
\coordinate (disttextPos) at (leftX |- dist.north);
\node[stepbox, fill=green!8, draw=green!35!black, anchor=north west] (disttext) at (disttextPos)
  {\textbf{Step 2: Analytical target}\\[2pt]
  $\displaystyle P(k;\phi)=\frac{1}{2^{2n_a}}
    \frac{\sin^2\!\left(\pi(\phi - k)\right)}
    {\sin^2\!\left(\dfrac{\pi(\phi - k)}{2^{n_a}}\right)}$\\[6pt]
  $\phi = 2^{n_a}|E'|\,\tau/(2\pi)$, \ $E' = E_{\mathrm{FCI}} - c_0 - E_{\mathrm{nuc}}$\\[2pt]
  {\small computed directly from $E'$, $n_a$, and $\tau$ --- no QPE circuit simulation required.}};

% robust vertical anchor: whichever is lower, the plot or its text box
\path let \p1=(dist.south), \p2=(disttext.south) in
  \pgfextra{\pgfmathparse{min(\y1,\y2)}\xdef\stepTwoBottomY{\pgfmathresult}};
\coordinate (yRefStep2) at (0,\stepTwoBottomY pt);
\coordinate (step2Bottom) at (dist |- yRefStep2);

% ================================================================
%  STEP 3: PARAMETERIZED VQC (text left, circuit right)
% ================================================================
\node (vqc) at ($(step2Bottom)+(0,-2.1)$) {
\scalebox{0.62}{
\begin{quantikz}[row sep=0.4cm, column sep=0.3cm]
\lstick{$q_0$} & \gate{R_Y(\theta_1)} & \ctrl{1}   & \qw        & \gate{R_Y(\theta_5)} & \meter{} \\
\lstick{$q_1$} & \gate{R_Y(\theta_2)} & \control{} & \ctrl{1}   & \gate{R_Y(\theta_6)} & \meter{} \\
\lstick{$q_2$} & \gate{R_Y(\theta_3)} & \ctrl{1}   & \control{} & \gate{R_Y(\theta_7)} & \meter{} \\
\lstick{$q_3$} & \gate{R_Y(\theta_4)} & \control{} & \qw        & \gate{R_Y(\theta_8)} & \meter{}
\end{quantikz}
}
};
\coordinate (vqctextPos) at (leftX |- vqc.north);
\node[stepbox, fill=violet!7, draw=violet!35, anchor=north west] (vqctext) at (vqctextPos)
  {\textbf{Step 3: Parameterized VQC}\\[2pt]
  $R_Y$-$CZ$ ansatz with linear entangler. The circuit is measured to obtain the output distribution $P_{\boldsymbol\theta}(k)$.};

% robust vertical anchor: whichever is lower, the circuit or its text box
\path let \p1=(vqc.south), \p2=(vqctext.south) in
  \pgfextra{\pgfmathparse{min(\y1,\y2)}\xdef\stepThreeBottomY{\pgfmathresult}};
\coordinate (yRefStep3) at (0,\stepThreeBottomY pt);
\coordinate (step3Bottom) at (vqc |- yRefStep3);

% ================================================================
%  STEP 4: CLASSICAL OPTIMIZER (text right, per request)
% ================================================================
\node[draw, rectangle, rounded corners, fill=orange!12, thick,
      minimum width=3.1cm, minimum height=1.6cm, align=center]
      (opt) at ($(step3Bottom)+(0,-1.9)$) {\textbf{Classical Optimizer}\\(COBYLA)\\[2pt] Update $\boldsymbol\theta$};
\coordinate (opttextPos) at (leftX |- opt.north);
\node[stepbox, fill=orange!10, draw=orange!45!black, anchor=north west] (opttext) at (opttextPos)
  {\textbf{Step 4: Optimize}\\[2pt]
  Compare $P_{\boldsymbol\theta}(k)$ against the target:\\[4pt]
  $\displaystyle \mathcal{L}(\boldsymbol\theta)=\sum_{k=0}^{2^{n_a}-1}
  \sqrt{\bigl[P_{\boldsymbol\theta}(k)-P_{\text{gt}}(k;\phi)\bigr]^2}$\\[6pt]
  COBYLA proposes a new $\boldsymbol\theta$, which is fed back into the circuit.};

% ================================================================
%  ARROWS: main vertical flow
% ================================================================
\draw[arr] (molecule.south) -- (dist.north)
    node[midway, right, lbl, xshift=2pt] {$E'$};

\draw[arr] (dist.south) -- (vqc.north)
    node[midway, right, lbl, xshift=2pt] {target\\$P_{\text{gt}}(k;\phi)$};

\draw[arr] (vqc.south) -- (opt.north)
    node[midway, right, lbl, xshift=2pt] {measure\\$P_{\boldsymbol\theta}(k)$};

% feedback loop: optimizer -> back up the right-hand side into the circuit
\coordinate (rowRef) at ($(opt.east)+(2.4,0)$);
\coordinate (fb1) at (rowRef);
\coordinate (fb2) at (fb1 |- vqc.east);
\draw[arr] (opt.east) -- (fb1) -- (fb2) -- (vqc.east)
    node[pos=0.5, right, lbl, xshift=2pt, rotate=90] {update $\boldsymbol\theta$};

\end{tikzpicture}

% ================================================================
%  CAPTION  (now a real \caption, so it auto-numbers as "Figure 1")
% ================================================================
\caption{Training workflow for the analytically-trained VQC surrogate.
\textbf{Step 1:} the molecular Hamiltonian is constructed classically and the reference ground-state energy $E_{\mathrm{FCI}}$ is obtained via Full Configuration Interaction.
\textbf{Step 2:} the shifted energy $E' = E_{\mathrm{FCI}} - c_0 - E_{\mathrm{nuc}}$, together with the ancilla count $n_a$ and evolution
time $\tau$, is substituted directly into the analytical Dirichlet kernel to obtain the
target QPE measurement distribution $P_{\text{gt}}(k;\phi)$ --- no simulation of the QPE
circuit is required. \textbf{Step 3:} a parameterized $R_Y$-$CZ$ ansatz (linear
entangler) is measured to obtain the output distribution $P_{\boldsymbol\theta}(k)$.
\textbf{Step 4:} $P_{\boldsymbol\theta}(k)$ is compared against the target via the cost
function $\mathcal{L}(\boldsymbol\theta)$, and the classical COBYLA optimizer proposes an
updated parameter vector $\boldsymbol\theta$, which is fed back into the ansatz. Steps 3--4
repeat until convergence.}
\label{fig:workflow}
\end{figure*}

\subsection{VQC Ansatz Architectures}
\label{subsec:ansatz}

We investigate two VQC ansatz architectures applied to the $n_a = 11$ 
ancilla qubit register. The two architectures differ in their rotation 
gate structure and in the entangler topologies that are investigated, as 
detailed below and summarized in the experimental workflow of 
Section~\ref{subsec:workflow}.

\subsubsection*{$R_Y$-$R_Z$-$CZ$ Ansatz}

The first ansatz follows the general rotation-entangler-rotation design 
pattern used for QPE surrogates in prior work~\cite{liu2024learnqpe}, 
consisting of $p$ repeated layers followed by a final rotation block. Each 
of the $p$ intermediate layers comprises:
\begin{enumerate}
    \item A layer of single-qubit $R_Y(\theta)$ rotations applied to every 
    qubit.
    \item A layer of single-qubit $R_Z(\theta)$ rotations applied to every 
    qubit.
    \item A layer of $CZ$ entangling gates arranged according to the chosen 
    entangler topology.
\end{enumerate}
After the $p$ entangling layers, a final rotation block of $R_Y$ and $R_Z$ 
gates is applied to every qubit without a subsequent entangling layer. For 
$n_a$ qubits and $p$ layers, this ansatz has $M = 2(p+1)n_a$ variational 
parameters in total. For the $R_Y$-$R_Z$-$CZ$ ansatz, we consider both the 
\textbf{linear entangler} and the \textbf{full entangler} topologies 
(described below), as the topology comparison constitutes the first 
stage of the experimental workflow.

\subsubsection*{$R_Y$-$CZ$ Ansatz}

The second ansatz is a reduced variant in which each layer contains only 
$R_Y(\theta)$ rotations followed by $CZ$ entangling gates, with a final 
$R_Y$ rotation block applied after the last entangling layer. This gives 
$M = (p+1)n_a$ variational parameters for $p$ layers and $n_a$ qubits, 
exactly half the parameter count of the $R_Y$-$R_Z$-$CZ$ ansatz at the 
same depth. The reduced gate count and circuit depth of this ansatz lowers 
noise accumulation on real hardware while retaining the entanglement 
structure necessary for learning the QPE distribution. For the $R_Y$-$CZ$ 
ansatz, only the \textbf{linear entangler} topology is considered, 
motivated by the conclusion of Stage 1 that the linear entangler is 
optimal for NISQ hardware.

\subsubsection*{Entangler Topologies}

Two entangler topologies are defined and used as follows:
\begin{itemize}
    \item \textbf{Linear entangler}: $CZ$ gates are applied between 
    nearest-neighbor qubit pairs $(q_0, q_1), (q_1, q_2), \ldots, 
    (q_{n_a-2}, q_{n_a-1})$, yielding a circuit depth of $\mathcal{O}(pn_a)$. 
    This topology is used for \textit{both} the $R_Y$-$R_Z$-$CZ$ and 
    $R_Y$-$CZ$ ansatze.
    \item \textbf{Full entangler}: $CZ$ gates are applied between all 
    $\binom{n_a}{2}$ pairs of qubits, yielding a circuit depth of 
    $\mathcal{O}(pn_a^2)$. This topology is used \textit{only} for the 
    $R_Y$-$R_Z$-$CZ$ ansatz in the topology comparison of Stage 1.
\end{itemize}
The motivation for restricting the full entangler to Stage 1 is that 
its $\mathcal{O}(pn_a^2)$ depth is expected to be detrimental on NISQ 
hardware~\cite{cerezo2021variational}, and this expectation is confirmed 
by the Stage 1 results, after which the full entangler is discarded from 
all further experiments.

\begin{figure*}[h]
\centering
\begin{tikzpicture}
 
% ============================================================
%  (b) RY-RZ-CZ ansatz, linear entangler, n_a = 4 qubits, p = 1 layer
% ============================================================
\node (circB) at (0,0) {%
\scalebox{0.85}{
\begin{quantikz}[row sep=0.2cm, column sep=0.4cm]
\lstick{$q_0$} & \gate{R_Y(\theta_{1,1})}\gategroup[4,steps=1,style={fill=green!60!black,fill opacity=0.3,rounded corners,inner xsep=3pt,inner ysep=4pt},label style={color=green!60!black}]{$R_Y$ layer} & \gate{R_Z(\theta_{1,5})}\gategroup[4,steps=1,style={fill=violet,fill opacity=0.3,rounded corners,inner xsep=3pt,inner ysep=4pt},label style={color=violet}]{$R_Z$ layer} & \ctrl{1}\gategroup[4,steps=2,style={fill=red,fill opacity=0.3,rounded corners,inner xsep=3pt,inner ysep=4pt},label style={color=red}]{$CZ$ layer}\gategroup[4,steps=4,style={dashed,draw=cyan!70!blue,thick,rounded corners,inner xsep=10pt,inner ysep=20pt},label style={color=cyan!70!blue}]{$p$ layers}   & \qw        & \gate{R_Y(\theta_{2,1})} & \gate{R_Z(\theta_{2,5})} & \meter{} \\
\lstick{$q_1$} & \gate{R_Y(\theta_{1,2})} & \gate{R_Z(\theta_{1,6})} & \control{} & \ctrl{1}   & \gate{R_Y(\theta_{2,2})} & \gate{R_Z(\theta_{2,6})} & \meter{} \\
\lstick{$q_2$} & \gate{R_Y(\theta_{1,3})} & \gate{R_Z(\theta_{1,7})} & \ctrl{1}   & \control{} & \gate{R_Y(\theta_{2,3})} & \gate{R_Z(\theta_{2,7})} & \meter{} \\
\lstick{$q_3$} & \gate{R_Y(\theta_{1,4})} & \gate{R_Z(\theta_{1,8})} & \control{} & \qw        & \gate{R_Y(\theta_{2,4})} & \gate{R_Z(\theta_{2,8})} & \meter{}
\end{quantikz}
}};
\node[above=0.15cm of circB] {\textbf{(b)}};
 
% ============================================================
%  (c) RY-CZ ansatz, linear entangler, n_a = 4 qubits, p = 1 layer
% ============================================================
\node (circC) [right=1.0cm of circB] {%
\scalebox{0.85}{
\begin{quantikz}[row sep=0.2cm, column sep=0.4cm]
\lstick{$q_0$} & \gate{R_Y(\theta_{1,1})}\gategroup[4,steps=1,style={fill=green!60!black,fill opacity=0.3,rounded corners,inner xsep=3pt,inner ysep=4pt},label style={color=green!60!black}]{$R_Y$ layer} & \ctrl{1}\gategroup[4,steps=2,style={fill=red,fill opacity=0.3,rounded corners,inner xsep=3pt,inner ysep=4pt},label style={color=red}]{$CZ$ layer}\gategroup[4,steps=3,style={dashed,draw=cyan!70!blue,thick,rounded corners,inner xsep=10pt,inner ysep=20pt},label style={color=cyan!70!blue}]{$p$ layers}   & \qw        & \gate{R_Y(\theta_{2,1})} & \meter{} \\
\lstick{$q_1$} & \gate{R_Y(\theta_{1,2})} & \control{} & \ctrl{1}   & \gate{R_Y(\theta_{2,2})} & \meter{} \\
\lstick{$q_2$} & \gate{R_Y(\theta_{1,3})} & \ctrl{1}   & \control{} & \gate{R_Y(\theta_{2,3})} & \meter{} \\
\lstick{$q_3$} & \gate{R_Y(\theta_{1,4})} & \control{} & \qw        & \gate{R_Y(\theta_{2,4})} & \meter{}
\end{quantikz}
}};
\node[above=0.15cm of circC] {\textbf{(c)}};
 
% ============================================================
%  invisible box fitting (b)+(c) so (a) can be centered above them
% ============================================================
\node[fit=(circB)(circC), inner sep=0pt] (rowBC) {};
 
% ============================================================
%  (a) RY-RZ-CZ ansatz, FULL entangler, n_a = 4 qubits, p = 1 layer
% ============================================================
\node (circA) [above=1.5cm of rowBC] {%
\scalebox{1.0}{
\begin{quantikz}[row sep=0.2cm, column sep=0.4cm]
\lstick{$q_0$} & \gate{R_Y(\theta_{1,1})}\gategroup[4,steps=1,style={fill=green!60!black,fill opacity=0.3,rounded corners,inner xsep=3pt,inner ysep=4pt},label style={color=green!60!black}]{$R_Y$ layer} & \gate{R_Z(\theta_{1,5})}\gategroup[4,steps=1,style={fill=violet,fill opacity=0.3,rounded corners,inner xsep=3pt,inner ysep=4pt},label style={color=violet}]{$R_Z$ layer} & \ctrl{1}\gategroup[4,steps=6,style={fill=red,fill opacity=0.3,rounded corners,inner xsep=3pt,inner ysep=4pt},label style={color=red}]{$CZ$ layer}\gategroup[4,steps=8,style={dashed,draw=cyan!70!blue,thick,rounded corners,inner xsep=10pt,inner ysep=20pt},label style={color=cyan!70!blue}]{$p$ layers} & \ctrl{2} & \ctrl{3} & \qw      & \qw      & \qw      & \gate{R_Y(\theta_{2,1})} & \gate{R_Z(\theta_{2,5})} & \meter{} \\
\lstick{$q_1$} & \gate{R_Y(\theta_{1,2})} & \gate{R_Z(\theta_{1,6})} & \control{} & \qw      & \qw      & \ctrl{1} & \ctrl{2} & \qw      & \gate{R_Y(\theta_{2,2})} & \gate{R_Z(\theta_{2,6})} & \meter{} \\
\lstick{$q_2$} & \gate{R_Y(\theta_{1,3})} & \gate{R_Z(\theta_{1,7})} & \qw      & \control{} & \qw      & \control{} & \qw      & \ctrl{1} & \gate{R_Y(\theta_{2,3})} & \gate{R_Z(\theta_{2,7})} & \meter{} \\
\lstick{$q_3$} & \gate{R_Y(\theta_{1,4})} & \gate{R_Z(\theta_{1,8})} & \qw      & \qw      & \control{} & \qw      & \control{} & \control{} & \gate{R_Y(\theta_{2,4})} & \gate{R_Z(\theta_{2,8})} & \meter{}
\end{quantikz}
}};
\node[above=0.15cm of circA] {\textbf{(a)}};

\end{tikzpicture}
\caption{VQC ansatz architectures used as QPE surrogates ($p=1$ layer, $n_a=4$ qubits shown for illustration). \textbf{(a)} $R_Y$-$R_Z$-$CZ$ ansatz with the \emph{full} entangler: each repeated layer applies $R_Y$ and $R_Z$ rotations to every qubit followed by a $CZ$ gate between every pair of qubits, ending in a final $R_Y$-$R_Z$ rotation block. \textbf{(b)} $R_Y$-$R_Z$-$CZ$ ansatz with the \emph{linear} entangler, where $CZ$ gates connect only nearest-neighbor qubits. \textbf{(c)} $R_Y$-$CZ$ ansatz with the linear entangler: identical entangling structure to (b) but with the $R_Z$ rotations omitted, halving the number of variational parameters per layer. Colored regions mark the $R_Y$ layer (green), $R_Z$ layer (violet), $CZ$ entangling layer (red), and the block of $p$ repeated layers (blue, dashed).}
\label{fig:ansatz_circuits}
\end{figure*}

\subsection{Experimental Workflow}
\label{subsec:workflow}

The VQC configuration is selected on H$_2$ in four sequential stages,
each building on the conclusions of the previous one, and the selected
configuration is then validated on the remaining molecules, as
summarized in Table~\ref{tab:configs}. At every stage, the choice is
made by comparing the hardware-executed VQC distribution with the
Dirichlet target using the four metrics of
Section~\ref{subsec:metric_eval}, together with the count of the
dominant bitstring.

\textbf{Stage 1 --- Entangler topology
($R_Y$-$R_Z$-$CZ$, $p=1$):}
We compare the linear and full entangler topologies for the
$R_Y$-$R_Z$-$CZ$ ansatz at $p = 1$ layer under ideal simulation,
noisy simulation, and real hardware execution with ideal- and
noisy-simulator-trained parameters, each with and without DD. The
superior topology is adopted for all subsequent stages.

\textbf{Stage 2 --- Number of layers
($R_Y$-$R_Z$-$CZ$, linear, $p \in \{1,\ldots,5\}$):}
Using the linear entangler identified in Stage 1, we vary the number
of layers from $p = 1$ to $p = 5$ for the $R_Y$-$R_Z$-$CZ$ ansatz,
under the same simulation and hardware conditions as in Stage 1, again
with and without DD. This stage identifies the optimal number of layers and establishes whether DD improves hardware performance. Since DD
degraded the performance of the linear-entangler circuits in Stages 1
and 2, all subsequent hardware executions are performed without DD.

\textbf{Stage 3 --- Ansatz
(linear, $p=1$, no DD):}
With the selected entangler and number of layers, we compare the
$R_Y$-$R_Z$-$CZ$ and $R_Y$-$CZ$ ans\"atze under ideal simulation, noisy
simulation, and real hardware execution with ideal- and
noisy-simulator-trained parameters, and select the preferred ansatz.

\textbf{Stage 4 --- Training environment
($R_Y$-$CZ$, linear, $p=1$, no DD):}
For the selected $R_Y$-$CZ$ ansatz, we compare the hardware performance
of parameters trained on the ideal simulator with that of parameters
trained on the noisy simulator, using the hardware executions of
Stage 3, and fix the training environment.

The framework differs from that of Liu \textit{et al.}~\cite{liu2024learnqpe}
in how the training target is obtained where the target is the
measurement histogram of the full QPE circuit executed on an ideal simulator with a finite
number of shots, so the controlled-unitary operations and the inverse QFT must be constructed
and simulated for every new problem. In the present framework, no part of the QPE circuit is
constructed. The target is evaluated exactly from the Dirichlet kernel, free of shot noise,
and its only inputs are a reference ground-state energy, the number of ancilla qubits $n_a$
and the evolution time $\tau$. The VQC therefore acts on the $n_a$ ancilla qubits alone, and
the molecule enters the pipeline only through the phase $\phi$. The Dirichlet target
reproduces the output of an ideal QPE only if $\phi$ is the phase of the true ground-state
eigenvalue of $\hat{H}'$. We therefore use $E_\text{FCI}$, which is the exact ground-state
energy within the chosen basis set, so that $P_\text{gt}(k;\phi)$ coincides with the ideal QPE
distribution. An approximate reference would shift $\phi$, and the surrogate would then
reproduce a different distribution from the one QPE would produce. FCI is tractable for the
minimal-basis molecules studied here, but its cost grows exponentially with system size
(Section~\ref{sec:conclusion}).

\textbf{Validation on further molecules:}
The configuration fixed in Stages 1--4, a single-layer linear-entangler
$R_Y$-$CZ$ ansatz with ideal-simulator-trained parameters executed
without DD, is trained for LiH, H$_2$O, BeH$_2$, NH$_3$, CH$_4$, O$_2$ and N$_2$ and executed on the ideal simulator and on real hardware.
For each molecule, the dominant bitstring is compared with that of the
Dirichlet target and converted into an energy via
Eq.~\eqref{eq:energy_recovery_correct}.

Two additional analyses are reported in the Supplementary Material. In
Section~\ref{sec:sidelobes} of the Supplementary Material, the side-lobe structure of the Stage~3 hardware
distributions is compared with the Dirichlet target, and in Section~\ref{sec:noise_analysis}, the
selected $R_Y$-$CZ$ ansatz with ideal-simulator-trained parameters is
executed on real hardware at $p \in \{8, 64\}$, with and without DD, to
characterize how the effectiveness of DD changes with circuit depth.

\begin{table}[h]
\centering
\caption{Summary of experimental configurations. ``Ideal Tr.'' and
``Noisy Tr.'' refer to the simulator training environment. ``Real HW''
indicates execution on IBM Quantum hardware. ``DD'' indicates whether
XpXm Dynamical Decoupling (as implemented in Qiskit) was applied during
hardware execution. Stage 4 uses the $R_Y$-$CZ$ hardware executions of
Stage 3. The rows marked App.~C correspond to the noise analysis in the
Supplementary Material.}
\label{tab:configs}
\renewcommand{\arraystretch}{1.3}
\resizebox{\linewidth}{!}{%
\begin{tabular}{clccccc}
\hline\hline
\textbf{Stage} & \textbf{Configuration} & \textbf{Ansatz} &
\textbf{Entangler} & \textbf{Layers} &
\textbf{Real HW} & \textbf{DD} \\
\hline
\multirow{4}{*}{1}
& Ideal Sim.            & $R_Y$-$R_Z$-$CZ$ & Linear \& Full & 1 & No  & -- \\
& Noisy Sim.            & $R_Y$-$R_Z$-$CZ$ & Linear \& Full & 1 & No  & -- \\
& Real + Noisy Tr.      & $R_Y$-$R_Z$-$CZ$ & Linear \& Full & 1 & Yes & No/Yes \\
& Real + Ideal Tr.      & $R_Y$-$R_Z$-$CZ$ & Linear \& Full & 1 & Yes & No/Yes \\
\hline
\multirow{4}{*}{2}
& Ideal Sim.            & $R_Y$-$R_Z$-$CZ$ & Linear & 1--5 & No  & -- \\
& Noisy Sim.            & $R_Y$-$R_Z$-$CZ$ & Linear & 1--5 & No  & -- \\
& Real + Noisy Tr.      & $R_Y$-$R_Z$-$CZ$ & Linear & 1--5 & Yes & No/Yes \\
& Real + Ideal Tr.      & $R_Y$-$R_Z$-$CZ$ & Linear & 1--5 & Yes & No/Yes \\
\hline
\multirow{4}{*}{3}
& Ideal Sim.            & $R_Y$-$R_Z$-$CZ$ / $R_Y$-$CZ$ & Linear & 1 & No  & -- \\
& Noisy Sim.            & $R_Y$-$R_Z$-$CZ$ / $R_Y$-$CZ$ & Linear & 1 & No  & -- \\
& Real + Noisy Tr.      & $R_Y$-$R_Z$-$CZ$ / $R_Y$-$CZ$ & Linear & 1 & Yes & No \\
& Real + Ideal Tr.      & $R_Y$-$R_Z$-$CZ$ / $R_Y$-$CZ$ & Linear & 1 & Yes & No \\
\hline
\multirow{2}{*}{4}
& Real + Noisy Tr.      & $R_Y$-$CZ$ & Linear & 1 & Yes & No \\
& Real + Ideal Tr.      & $R_Y$-$CZ$ & Linear & 1 & Yes & No \\
\hline
\multirow{2}{*}{Valid.}
& Ideal Sim.            & $R_Y$-$CZ$ & Linear & 1 & No  & -- \\
& Real + Ideal Tr.      & $R_Y$-$CZ$ & Linear & 1 & Yes & No \\
\hline
\multirow{2}{*}{App.~C}
& Real + Ideal Tr.      & $R_Y$-$CZ$ & Linear & 8, 64 & Yes & No \\
& Real + Ideal Tr. + DD & $R_Y$-$CZ$ & Linear & 8, 64 & Yes & Yes \\
\hline\hline
\end{tabular}}
\end{table}

\subsection{Training Protocol}
\label{subsec:training}
The VQC parameters $\boldsymbol{\theta} \in \mathbb{R}^M$ are optimized 
using the COBYLA optimizer~\cite{powell1994direct} to minimize the cost 
function
\begin{equation}
    \mathcal{L}(\boldsymbol{\theta}) = \sum_{k=0}^{2^{n_a}-1}
    \sqrt{\bigl[P_{\boldsymbol{\theta}}(k) - P_\text{gt}(k;\phi)\bigr]^2},
    \label{eq:cost_main}
\end{equation}
the $L^1$ distance between the VQC output distribution and the exact
analytical target (see Section~\ref{subsec:cost} of the Supplementary Material),
where $M$ is the total 
number of variational parameters of the chosen ansatz ($M = 2(p+1)n_a$ 
for the $R_Y$-$R_Z$-$CZ$ ansatz, $M = (p+1)n_a$ for the $R_Y$-$CZ$ 
ansatz, as defined in Section~\ref{subsec:ansatz}). The ansatz 
parameters are initialized to zero ($\boldsymbol{\theta}_0 = \mathbf{0}$) 
at the start of each optimization run. At each COBYLA iteration, the VQC 
output distribution $P_{\boldsymbol{\theta}}(k)$ is estimated by 
executing the parameterized circuit on the appropriate simulator 
backend with $N_\text{shots} = 100{,}000$ shots, as
$P_{\boldsymbol{\theta}}(k) = c_k/N_\text{shots}$, where $c_k$ is the
count of bitstring $k$. The target $P_\text{gt}(k;\phi)$ in
Eq.~\eqref{eq:cost_main} is used as the exact probability vector and is
not sampled.

We perform two distinct training modes:
\begin{itemize}
    \item \textbf{Ideal simulator training}: The VQC circuit is 
    executed on the \texttt{AerSimulator} backend with no noise model, 
    providing shot-based probability estimates under ideal noiseless 
    conditions~\cite{qiskit2023}.
    \item \textbf{Noisy simulator training}: A noise model is 
    extracted directly from the calibrated \texttt{ibm\_boston} 
    superconducting device via the \texttt{QiskitRuntimeService} 
    using \texttt{NoiseModel.from\_backend()}~\cite{qiskit2023}. 
    This noise model, which incorporates the device-specific gate 
    error rates, readout errors, decoherence parameters, and basis 
    gate set of \texttt{ibm\_boston}, is used to configure an 
    \texttt{AerSimulator} instance (\texttt{ibm\_boston\_sim}) that 
    faithfully emulates the noise environment of the real hardware 
    during training. The VQC circuit is transpiled to the native 
    basis gates of \texttt{ibm\_boston} before each 
    execution~\cite{preskill2018quantum, endo2021hybrid, qiskit2023}.
\end{itemize}

In both training modes, the circuit is transpiled using 
\texttt{Qiskit}'s \texttt{transpile} function with 
\texttt{optimization\_level=3} targeting the respective 
backend~\cite{qiskit2023}, and the COBYLA optimizer is run with 
a maximum iteration budget of $4 \times 10^7$, which is found to 
be more than sufficient for convergence across all configurations 
tested.

\subsection{Real Hardware Execution}
\label{subsec:hardware}

All real hardware experiments are executed on the IBM Quantum
superconducting device \texttt{ibm\_boston}, accessed via the
\texttt{Qiskit} SDK~\cite{qiskit2023}. This is the same device whose
calibration data define the noise model used for noisy-simulator
training (Section~\ref{subsec:training}), so that noisy-simulator-trained
parameters are evaluated on the hardware they were trained to emulate. The
trained VQC parameters $\boldsymbol{\theta}^*$, obtained from either 
ideal or noisy classical simulation, are transferred directly to the 
hardware circuit without further optimization. This zero-shot transfer 
protocol tests the generalization of the learned parameters to the 
real device noise environment. Prior to execution, each circuit is 
transpiled to the native gate set of the target device using 
\texttt{Qiskit}'s transpiler with optimization level 
3~\cite{qiskit2023}. Each hardware experiment uses 
$N_\text{shots} = 100{,}000$ measurement shots, consistent with the 
shot budget used during training.

\subsection{Dynamical Decoupling Implementation}
\label{subsec:dd_impl}

For configurations incorporating Dynamical Decoupling (DD), we apply 
the XpXm pulse sequence, as implemented in Qiskit's dynamical-decoupling 
pass, to all idle qubits during circuit execution. The XpXm sequence 
alternates $\pi$-pulses about the $+X$ and $-X$ axes, providing 
first-order suppression of dephasing noise during idle windows. DD 
sequences are inserted into the transpiled circuit using \texttt{Qiskit}'s 
\texttt{PadDynamicalDecoupling} pass applied to the dynamically 
scheduled circuit~\cite{qiskit2023}. The DD pulses are placed 
symmetrically within each identified idle window subject to the 
hardware timing constraints. The depth-dependent effectiveness of DD 
is examined systematically in Section~\ref{sec:noise_analysis} of the Supplementary Material.

\subsection{Metric Evaluation}
\label{subsec:metric_eval}

Following hardware or simulation execution, the measurement histogram 
$P_{\boldsymbol{\theta}}(k)$ is compared against the analytical 
Dirichlet distribution $P_\text{gt}(k;\phi)$ using the four 
metrics defined in Section~\ref{subsec:metrics} of the Supplementary Material, which are Hellinger distance 
$H$, fidelity error $\epsilon_F$, total variation distance TVD, and 
normalized Jensen-Shannon divergence $\sqrt{\mathrm{JSD}/\ln 2}$. 
All metrics are evaluated over the full support of $2^{11} = 2048$ 
bitstrings, with a Laplace smoothing correction 
$\epsilon_\text{smooth} = 10^{-10}$ applied to zero-count bins to ensure numerically stable evaluation of the logarithmic terms in the JSD computation.

For the dominant bitstring analysis, we additionally report the count 
of the dominant bitstring of the Dirichlet target, $k^* =
\texttt{00101010011}$ for H$_2$, across all experimental configurations,
which serves as a direct indicator of how well the hardware-executed VQC
concentrates probability mass at the correct QPE outcome. The dominant
bitstrings of the remaining molecules are listed in Table~A1 of the
Supplementary Material. The recovered ground-state energy is 
computed from $k^*$ via Eq.~\eqref{eq:energy_recovery_correct} and 
compared against $E_\text{FCI}$ via Eq.~\eqref{eq:energy_error} to assess whether the output lies 
within chemical accuracy 
($|\Delta E| < 1.59 \times 10^{-3}$~Hartree)~\cite{mcardle2020quantum, 
helgaker2000molecular}.

\subsection{Software and Hardware Stack}
\label{subsec:software}

All simulations and circuit constructions are implemented in Python 
using \texttt{Qiskit}~\cite{qiskit2023} for circuit definition, 
transpilation, and hardware execution; \texttt{Qiskit 
Nature}~\cite{qiskit2023} for Hamiltonian construction and the
Jordan--Wigner mapping; \texttt{PySCF}~\cite{sun2018pyscf} for molecular integral 
computation and FCI energy evaluation; \texttt{NumPy}~\cite{harris2020array} 
and \texttt{SciPy}~\cite{virtanen2020scipy} for numerical evaluation 
of the Dirichlet kernel, metric computation, and post-processing. 
Visualization is performed using 
\texttt{Matplotlib}~\cite{hunter2007matplotlib}.

\section{Results and Discussion}
\label{sec:results}

In this section, we present the results of the stepwise selection
described in Section~\ref{subsec:workflow} and of the validation on
further molecules. We first verify that the VQC can be trained against
the analytical Dirichlet target
(Section~\ref{subsec:results_validation}). We then report the four
selection stages on H$_2$: the entangler topology
(Section~\ref{subsec:results_topology}), the number of layers together
with the effect of dynamical decoupling (Section~\ref{subsec:dd_layers}),
the ansatz (Section~\ref{subsec:results_ansatz}), and the training
environment (Section~\ref{subsec:results_training}). Finally, we discuss
the recovered H$_2$ energy (Section~\ref{subsec:results_energy}),
summarize the selection results (Section~\ref{subsec:results_summary}),
and validate the selected configuration on seven further molecules
(Section~\ref{subsec:results_molecules}).

\subsection{Training Against the Analytical Dirichlet Target}
\label{subsec:results_validation}

\begin{figure}[htbp]
    \centering
    \includegraphics[width=0.8\linewidth]{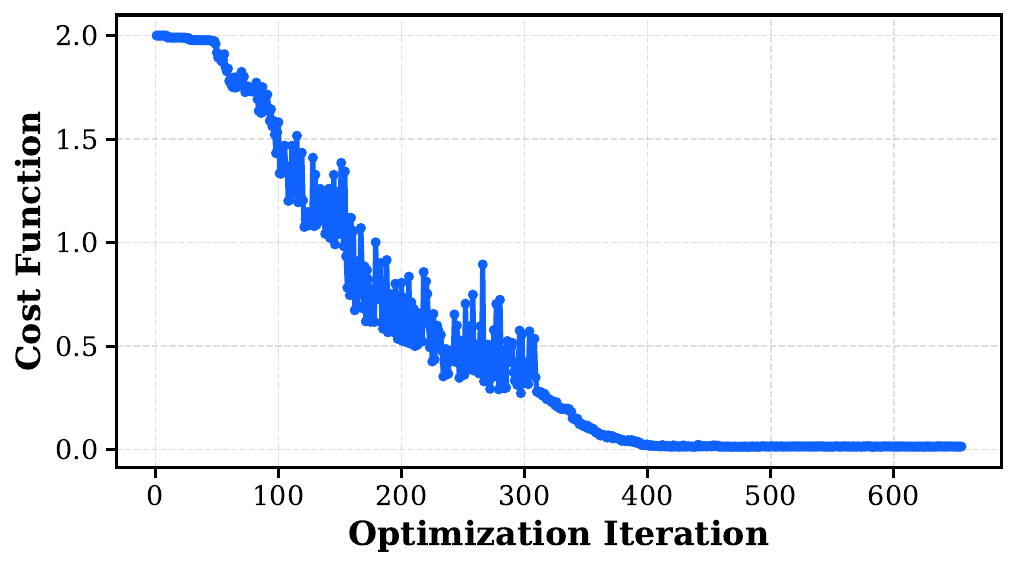}
    \caption{Convergence of the cost function
    $\mathcal{L}(\boldsymbol{\theta})$ during COBYLA optimization
    of the $R_Y$-$R_Z$-$CZ$ VQC (linear entangler, $p=1$) trained
    on the analytical Dirichlet distribution for H$_2$.}
    \label{fig:dirichlet_vqc_convergence}
\end{figure}

We first verify that a VQC trained on the analytical Dirichlet
distribution learns the QPE output distribution of H$_2$.
Figure~\ref{fig:dirichlet_vqc_convergence} shows the cost function
$\mathcal{L}(\boldsymbol{\theta})$ over the course of COBYLA optimization
for the $R_Y$-$R_Z$-$CZ$ ansatz with linear entangler at $p = 1$. The cost
decreases and converges to a well-defined minimum, showing that the
analytical target provides a usable training signal for the variational
optimization~\cite{liu2024learnqpe, cerezo2021variational}.

\subsection{Stage 1: Entangler Topology}
\label{subsec:results_topology}

\begin{figure}[htbp]
    \centering
    \includegraphics[width=\linewidth]{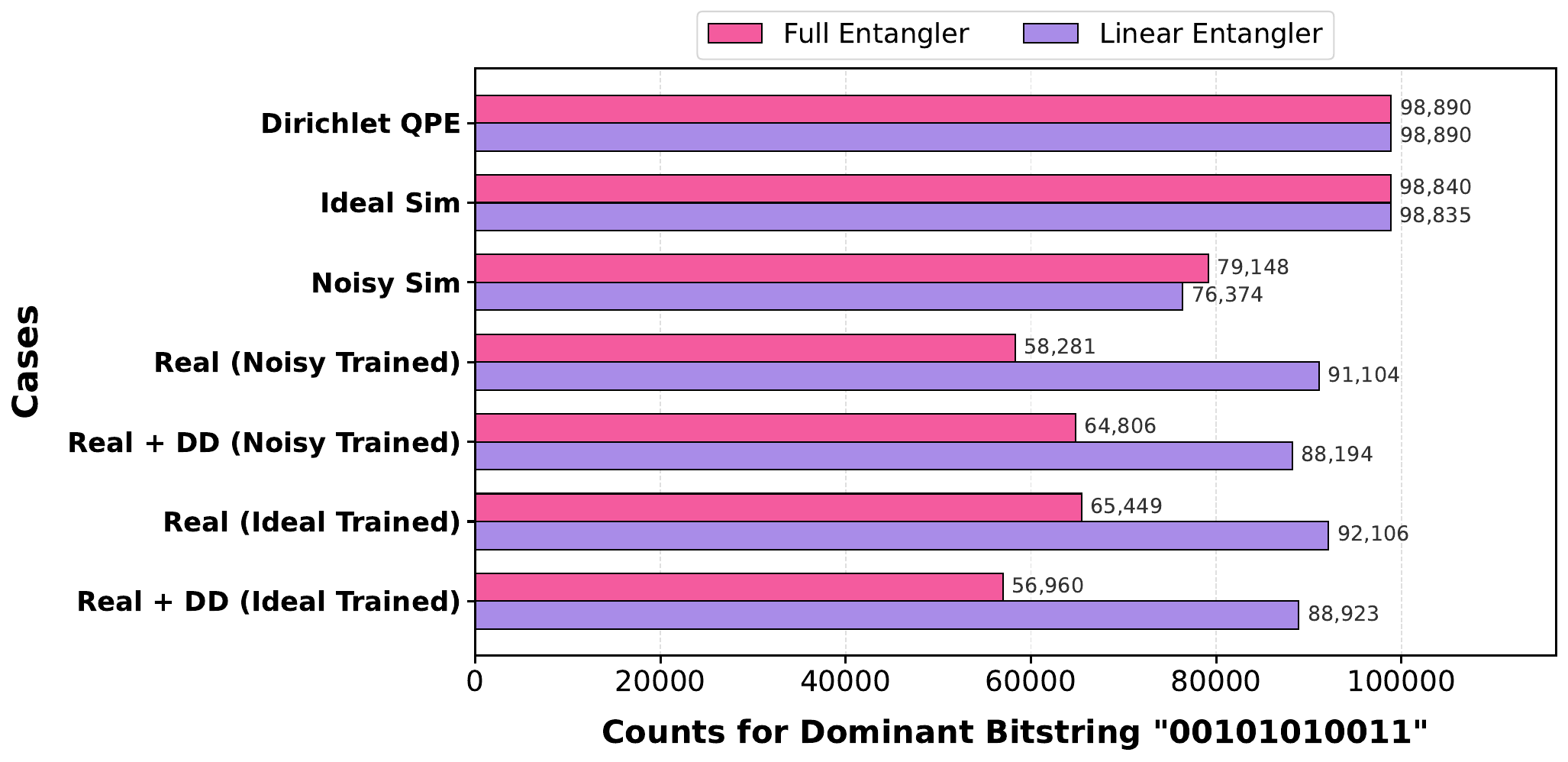}
    \caption{Counts of the dominant bitstring \texttt{00101010011} for
    the $R_Y$-$R_Z$-$CZ$ ansatz at $p=1$, comparing the full entangler
    (pink) and the linear entangler (purple) across all experimental
    cases ($N_\text{shots} = 100{,}000$). Real-hardware cases are
    labeled by the simulator used for training (noisy or ideal) and by
    whether DD was applied. On real hardware, the linear entangler
    achieves substantially higher counts than the full entangler in
    every case.}
    \label{fig:topology_dominant_bitstring}
\end{figure}

The first stage compares the linear and full entangler topologies for the
$R_Y$-$R_Z$-$CZ$ ansatz at $p = 1$.
Figure~\ref{fig:topology_dominant_bitstring} shows the counts of the
dominant bitstring \texttt{00101010011} across all experimental cases for
both topologies.

\begin{figure}[htbp]
    \centering
    \includegraphics[width=\linewidth]{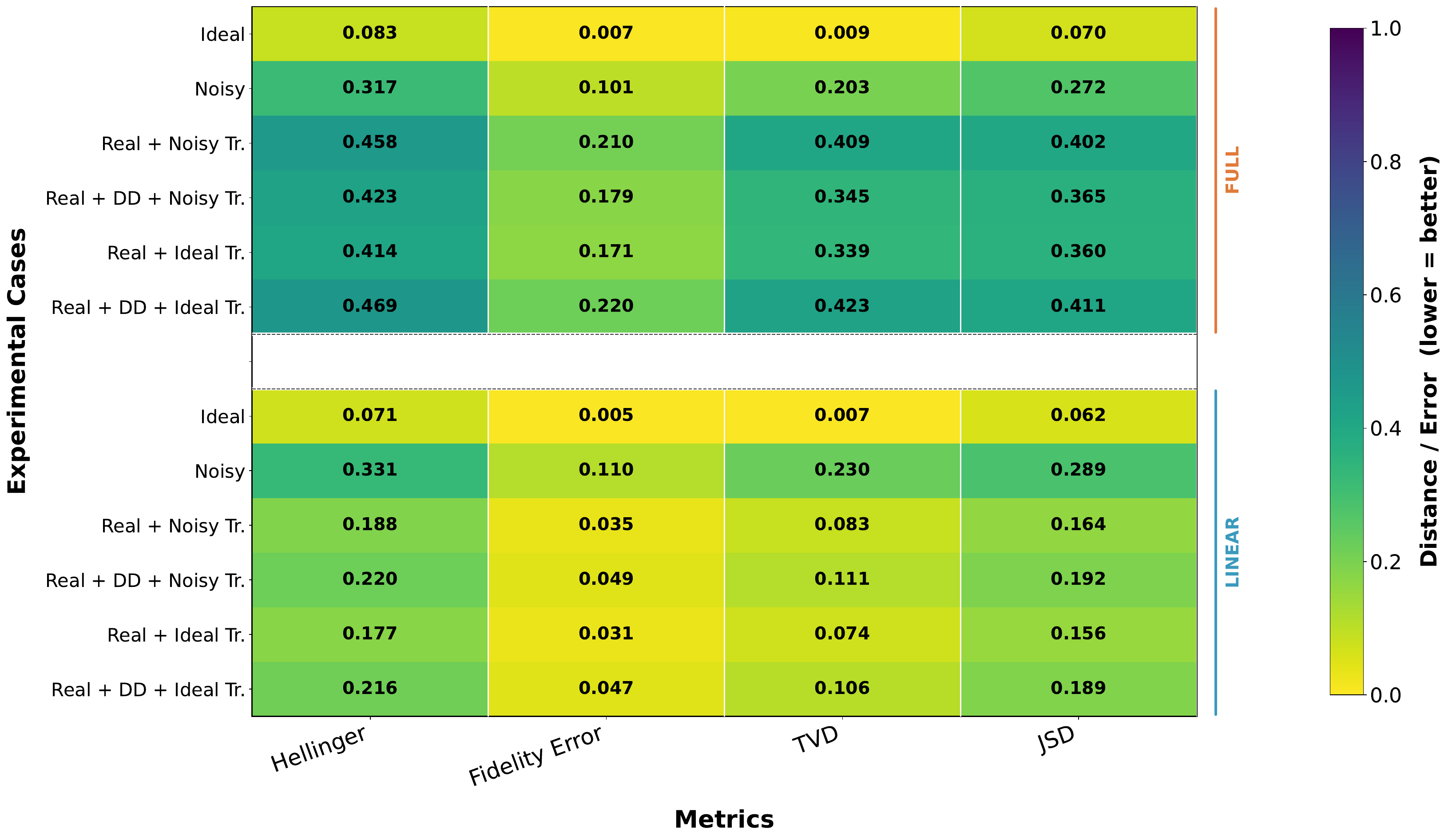}
    \caption{Metric heatmap (Hellinger distance, fidelity error, TVD,
    JSD) comparing the $R_Y$-$R_Z$-$CZ$ ansatz with full entangler
    (top, orange) and linear entangler (bottom, blue) at $p=1$,
    across all experimental cases. Lower values indicate better
    agreement with the Dirichlet target. The linear entangler
    achieves lower metric values in all hardware configurations.}
    \label{fig:heatmap_full_linear}
\end{figure}

Under ideal and noisy simulation, both topologies achieve comparable
dominant-bitstring counts (full: $98{,}840$ and $79{,}148$; linear:
$98{,}835$ and $76{,}374$, respectively), indicating that the
difference in expressibility between the two topologies is minimal for
this learning task at $p=1$. On real hardware, however, the gap becomes
large: the linear entangler achieves $91{,}104$ counts with
noisy-trained parameters and $92{,}106$ counts with ideal-trained
parameters, whereas the full entangler achieves only $58{,}281$ and
$65{,}449$, respectively. This difference reflects the larger two-qubit
gate count of the full entangler, $n_a(n_a-1)/2 = 55$ $CZ$ gates per
layer compared with $n_a - 1 = 10$ for the linear entangler, which is
further increased on hardware with limited qubit connectivity by the
SWAP operations required to route non-neighboring
interactions~\cite{preskill2018quantum, bharti2022noisy}. The four
distributional metrics in Figure~\ref{fig:heatmap_full_linear} confirm
this picture. For the linear entangler, applying DD lowers the dominant
count, from $92{,}106$ to $88{,}923$ with ideal-trained parameters and
from $91{,}104$ to $88{,}194$ with noisy-trained parameters. For the full
entangler, the effect of DD is mixed: it raises the dominant count with
noisy-trained parameters ($58{,}281 \to 64{,}806$) but lowers it with
ideal-trained parameters ($65{,}449 \to 56{,}960$). In all cases, the
full-entangler counts remain far below those of the linear entangler.
Based on this analysis, the linear entangler is adopted for all
subsequent stages.

\subsection{Stage 2: Number of Layers and Effect of Dynamical Decoupling}
\label{subsec:dd_layers}

\begin{figure}[htbp]
    \centering
    \includegraphics[width=0.85\linewidth]{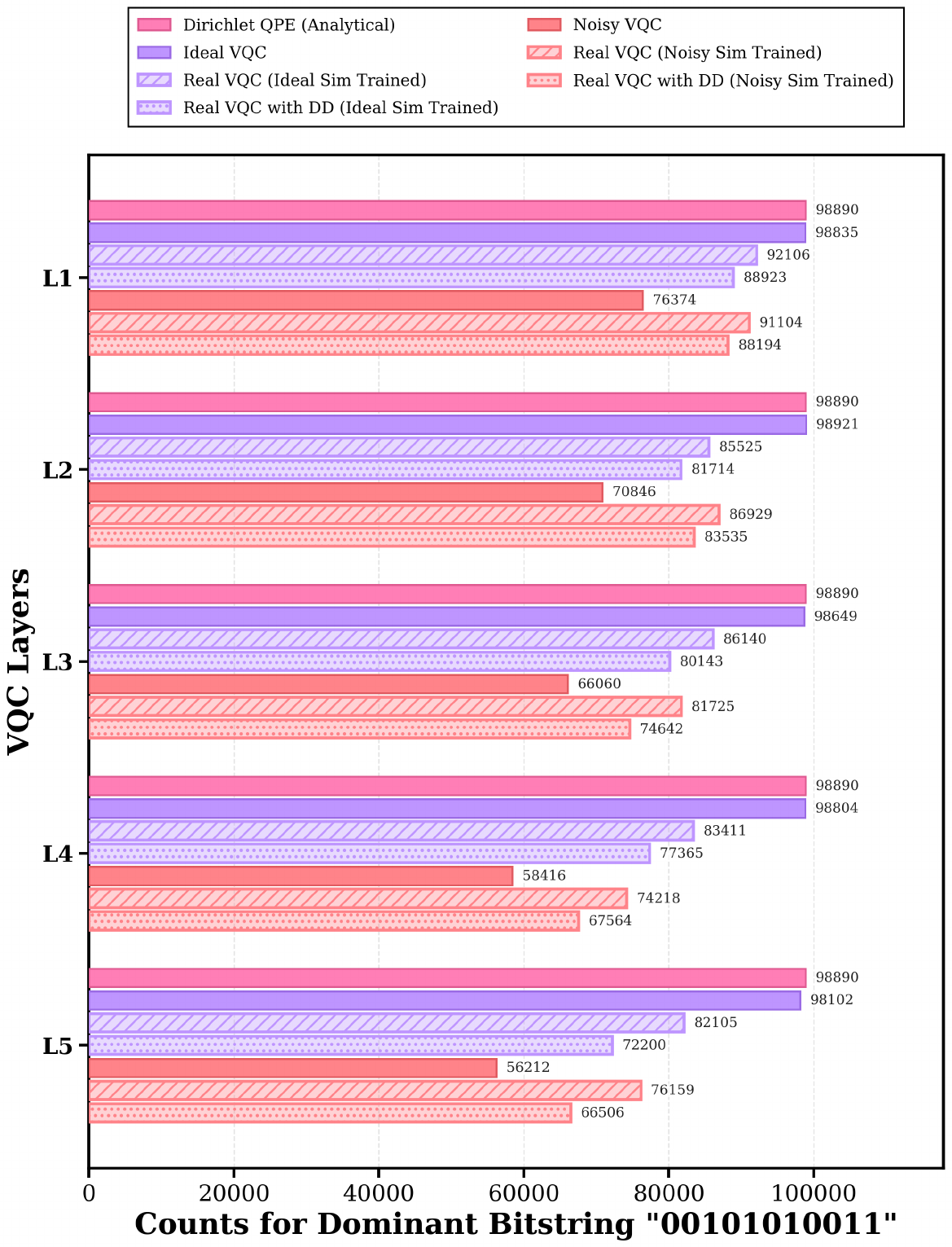}
    \caption{Counts of the dominant bitstring \texttt{00101010011} for
    the $R_Y$-$R_Z$-$CZ$ ansatz with linear entangler for
    $p \in \{1,\ldots,5\}$, comparing the analytical Dirichlet
    reference, ideal simulation, noisy simulation, real hardware with
    ideal- and noisy-simulator-trained parameters, and their
    DD-augmented variants ($N_\text{shots} = 100{,}000$). The ideal
    simulation closely tracks the Dirichlet reference at all depths. On
    real hardware, the counts generally decrease with increasing depth,
    and the non-DD circuits outperform their DD-augmented counterparts
    at every layer count.}
    \label{fig:dominant_bitstring_layers}
\end{figure}

\begin{figure}[htbp]
    \centering
    \includegraphics[width=\linewidth]{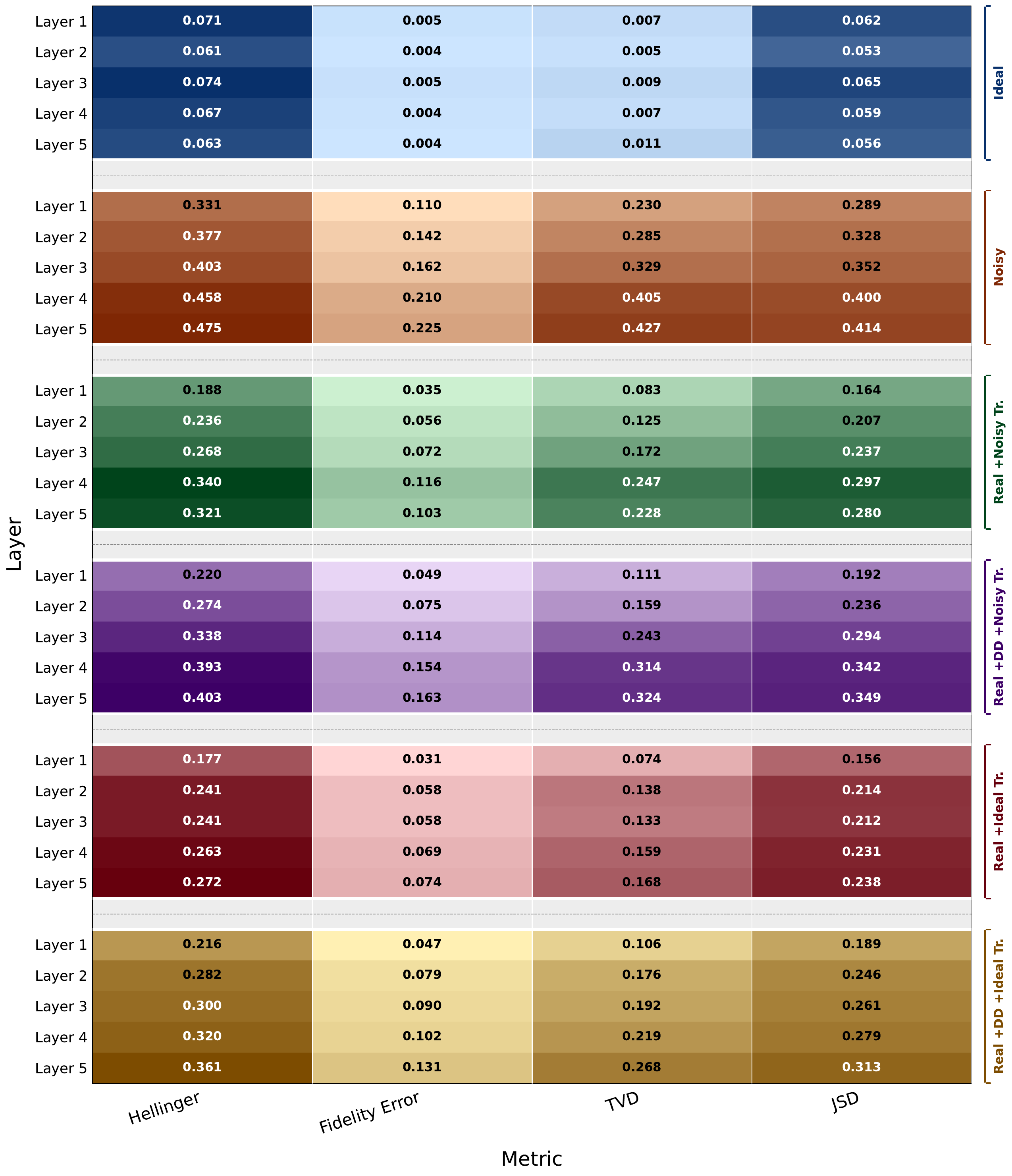}
    \caption{Metric heatmap (Hellinger distance, fidelity error, TVD,
    JSD) for the $R_Y$-$R_Z$-$CZ$ ansatz with linear entangler for each
    $p \in \{1,\ldots,5\}$ under all experimental cases. Each colored
    block represents a distinct experimental case. Lower metric values
    indicate better agreement with the Dirichlet target. On hardware,
    the metrics generally worsen with increasing depth, and $p = 1$
    achieves the best performance.}
    \label{fig:heatmap_layers}
\end{figure}

With the linear entangler fixed, we vary the number of layers from
$p = 1$ to $p = 5$ for the $R_Y$-$R_Z$-$CZ$ ansatz under all
experimental cases, with and without DD.
Figure~\ref{fig:dominant_bitstring_layers} shows the dominant-bitstring
counts, and Figure~\ref{fig:heatmap_layers} the corresponding metrics.

\textbf{Ideal simulation.} Under ideal conditions, all depths achieve
low metric values (Hellinger $\approx 0.061$--$0.074$, fidelity error
$\approx 0.004$--$0.005$), with no significant trend across layers, and
the dominant count remains between $98{,}102$ and $98{,}921$. The ansatz
therefore learns the Dirichlet distribution at all depths in the absence
of noise.

\textbf{Noisy simulation and hardware.} Under noisy simulation, the
metrics degrade monotonically with depth (Hellinger increases from
$0.331$ at $p=1$ to $0.475$ at $p=5$), and the dominant count decreases
from $76{,}374$ to $56{,}212$, reflecting noise accumulation from the
additional two-qubit gates. A similar overall decline is observed on real
hardware, where the dominant count decreases from $92{,}106$ at $p=1$ to
$82{,}105$ at $p=5$ with ideal-trained parameters and from $91{,}104$ to
$76{,}159$ with noisy-trained parameters, although the decrease is not
strictly monotonic between adjacent depths.

\textbf{Effect of dynamical decoupling.} At every layer count and for
both training environments, the DD-augmented hardware executions
underperform their non-DD counterparts, and the gap widens with depth.
With ideal-trained parameters, DD lowers the dominant count by
$3{,}183$ at $p=1$ ($92{,}106 \to 88{,}923$) and by $9{,}905$ at $p=5$
($82{,}105 \to 72{,}200$); with noisy-trained parameters, the reduction
grows from $2{,}910$ to $9{,}653$. The metrics show the same behavior:
with noisy-trained parameters, the Hellinger distance is $0.188$ without
DD and $0.220$ with DD at $p=1$, and $0.321$ versus $0.403$ at $p=5$.
Because DD degraded the performance of the linear-entangler circuits
at every tested depth, in both Stage 1 and Stage 2, all subsequent
hardware executions are performed without DD. The dependence
of DD on circuit depth is examined further in Section~\ref{sec:noise_analysis} of the
Supplementary Material.

\textbf{Selected number of layers.} The single-layer circuit $p=1$
achieves the best hardware performance for both training environments
and is adopted for all subsequent stages.

\subsection{Stage 3: Ansatz}
\label{subsec:results_ansatz}

\begin{figure}[htbp]
    \centering
    \includegraphics[width=\linewidth]{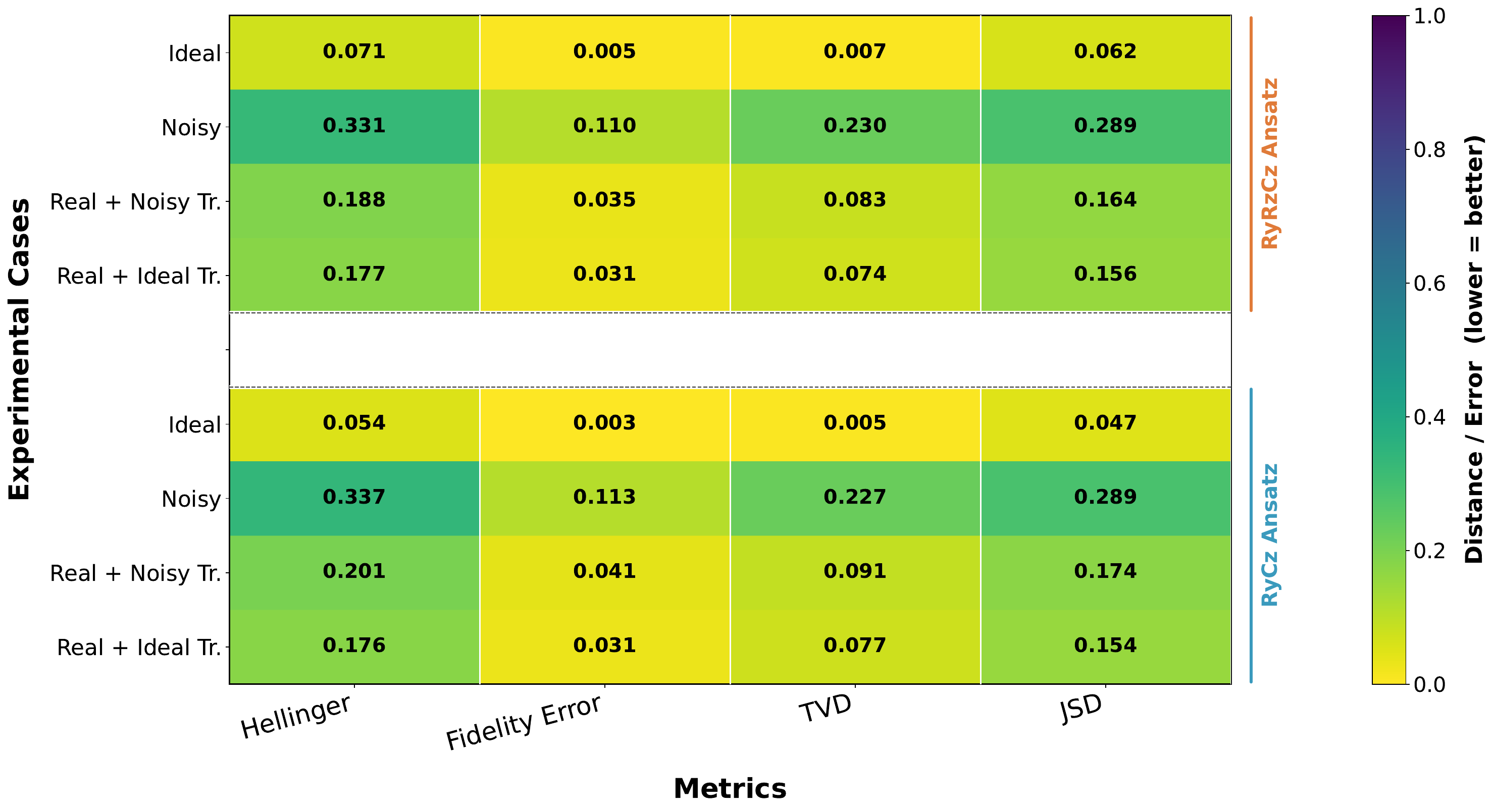}
    \caption{Metric comparison (Hellinger distance, fidelity error,
    TVD, JSD) between the $R_Y$-$R_Z$-$CZ$ ansatz (top, orange)
    and the $R_Y$-$CZ$ ansatz (bottom, blue) with linear entangler
    at $p=1$, evaluated under ideal simulation, noisy simulation, real
    hardware with noisy-simulator-trained parameters
    (Real + Noisy Tr.), and real hardware with ideal-simulator-trained
    parameters (Real + Ideal Tr.). Both ans\"atze achieve comparable
    metric values in all cases.}
    \label{fig:heatmap_ansatz_comparison}
\end{figure}

With the linear entangler and $p = 1$ fixed and DD omitted, Stage 3
compares the $R_Y$-$R_Z$-$CZ$ ansatz with the reduced $R_Y$-$CZ$ ansatz.
Figure~\ref{fig:heatmap_ansatz_comparison} shows the metrics of both
ans\"atze across the four non-DD experimental cases.

The two ans\"atze achieve closely comparable metric values in all
experimental cases. Under ideal simulation, the $R_Y$-$CZ$ ansatz is
marginally better (Hellinger $0.054$ vs.\ $0.071$). On real hardware with
ideal-trained parameters, the two are nearly indistinguishable
(Hellinger $0.176$ vs.\ $0.177$, fidelity error $0.031$ for both, TVD
$0.077$ vs.\ $0.074$, JSD $0.154$ vs.\ $0.156$ for $R_Y$-$CZ$ and
$R_Y$-$R_Z$-$CZ$, respectively), and their dominant-bitstring counts
differ by less than $1\%$ ($91{,}499$ vs.\ $92{,}106$). The additional
$R_Z$ rotations therefore provide no meaningful advantage for learning
the sharply peaked Dirichlet distribution of H$_2$, in which the
dominant bitstring carries $98.9\%$ of the probability mass. Since the
$R_Y$-$CZ$ ansatz achieves equivalent accuracy with half the rotation
parameters and fewer gates, it is selected for the remainder of this
work.

\subsection{Stage 4: Training Environment}
\label{subsec:results_training}

\begin{figure}[htbp]
    \centering
    \includegraphics[width=\linewidth]{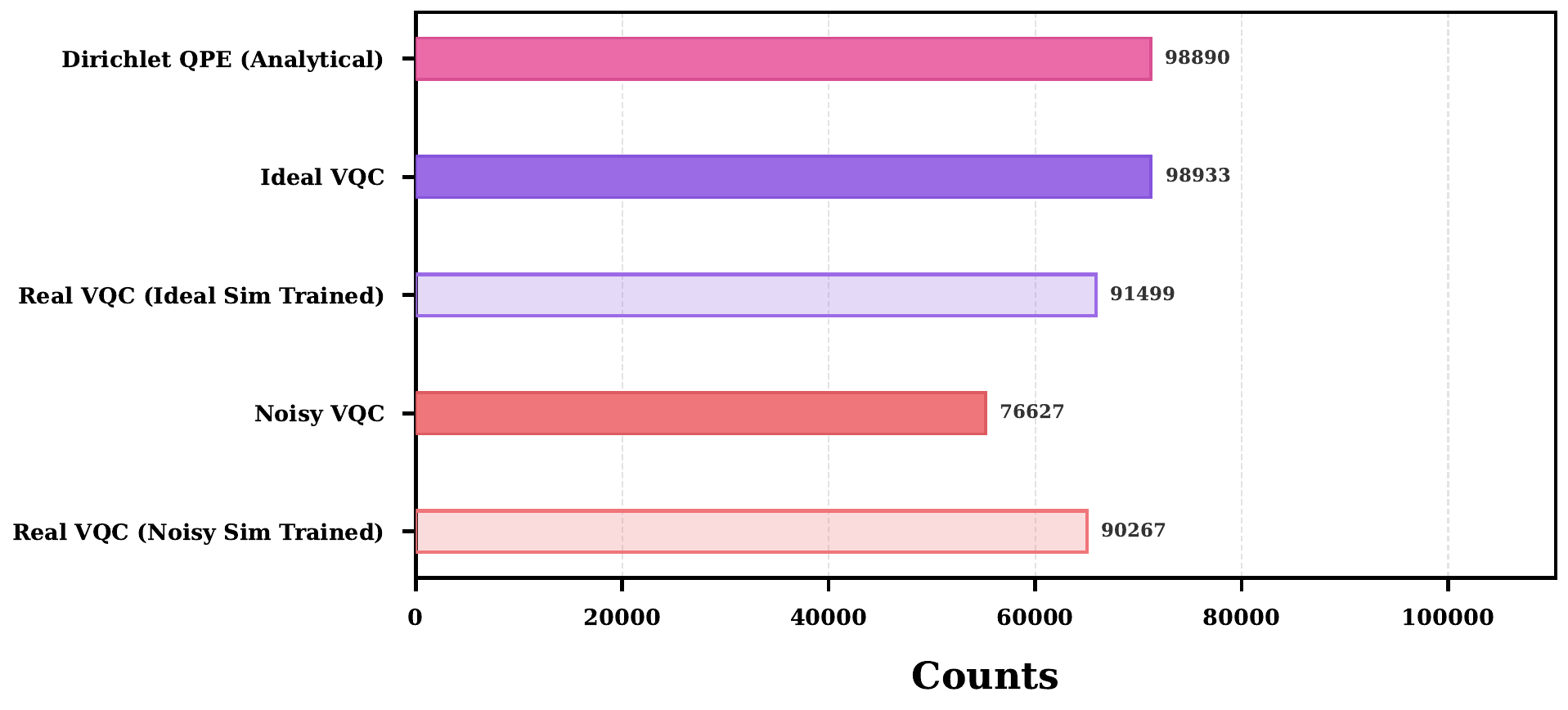}
    \caption{Counts of the dominant bitstring \texttt{00101010011} for
    the $R_Y$-$CZ$ ansatz with linear entangler at $p=1$: analytical
    Dirichlet reference ($98{,}890$), ideal simulation ($98{,}933$),
    noisy simulation ($76{,}627$), real hardware with ideal-trained
    parameters ($91{,}499$), and real hardware with noisy-trained
    parameters ($90{,}267$).}
    \label{fig:dominant_bitstring_rycz}
\end{figure}

\begin{figure}[htbp]
    \centering
    \includegraphics[width=\linewidth]{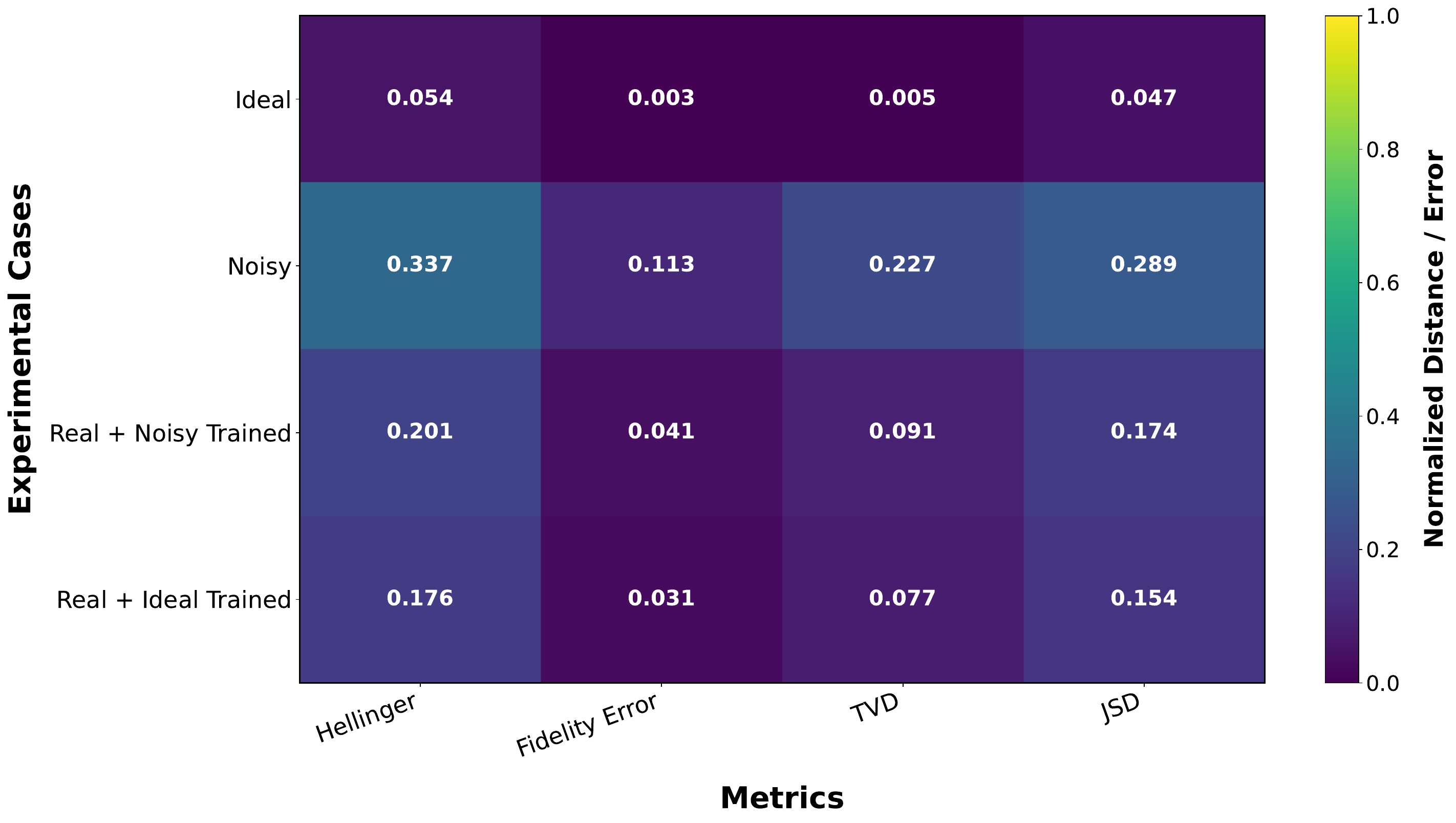}
    \caption{Metric heatmap (Hellinger distance, fidelity error, TVD,
    JSD) for the $R_Y$-$CZ$ ansatz with linear entangler at $p=1$,
    evaluated under ideal simulation, noisy simulation, real hardware
    with noisy-simulator-trained parameters, and real hardware with
    ideal-simulator-trained parameters. Among the hardware executions,
    the ideal-trained configuration achieves the lowest metric values
    (Hellinger $= 0.176$, fidelity error $= 0.031$, TVD $= 0.077$,
    JSD $= 0.154$).}
    \label{fig:heatmap_rycz}
\end{figure}

For the selected $R_Y$-$CZ$ ansatz, the final stage compares parameters
trained on the ideal simulator with parameters trained on the noisy
simulator, both executed on \texttt{ibm\_boston}.
Figure~\ref{fig:dominant_bitstring_rycz} shows the dominant-bitstring
counts and Figure~\ref{fig:heatmap_rycz} the metrics.

On real hardware, the ideal-trained $R_Y$-$CZ$ circuit achieves
$91{,}499$ counts on the dominant bitstring, compared with $90{,}267$ for
the noisy-trained circuit. Although the difference in counts is modest,
the ideal-trained parameters perform better on all four metrics
(Hellinger $0.176$ vs.\ $0.201$, fidelity error $0.031$ vs.\ $0.041$, TVD
$0.077$ vs.\ $0.091$, JSD $0.154$ vs.\ $0.174$). The same ordering holds
for the $R_Y$-$R_Z$-$CZ$ ansatz (Figure~\ref{fig:heatmap_ansatz_comparison}). A
possible explanation is that the noiseless cost landscape of ideal
training is easier to optimize, while the device noise model does not
reproduce the hardware closely: the noisy simulator yields
substantially worse metrics than the hardware itself (Hellinger $0.337$
vs.\ $0.176$--$0.201$ for $R_Y$-$CZ$), indicating that it overestimates
the noise experienced by these shallow
circuits~\cite{cerezo2021variational, bharti2022noisy}. Ideal-simulator
training is therefore adopted, which completes the selected
configuration: a single-layer linear-entangler $R_Y$-$CZ$ ansatz with
ideal-simulator-trained parameters, executed without DD.

\subsection{Energy Recovery for H$_2$}
\label{subsec:results_energy}

The figure of merit for molecular ground-state energy estimation is the
absolute error between the recovered energy and the FCI reference. For
H$_2$, the dominant bitstring $k^* = \texttt{00101010011}$ ($k^* = 339$)
yields the phase estimate $\theta^* = k^*/2^{n_a}$, from which the total
energy is recovered via Eq.~\eqref{eq:energy_recovery_correct}.
Table~\ref{tab:energy_results} lists the quantities entering this
calculation.

\begin{table}[h]
\centering
\caption{Ground-state energy recovery for H$_2$ (STO-3G, $R = 0.74$~\AA,
$n_a = 11$, $\tau = 1.0$~a.u.). All energies in Hartree. The recovered
energy is common to all H$_2$ configurations whose dominant bitstring is
$k^* = \texttt{00101010011}$.}
\label{tab:energy_results}
\renewcommand{\arraystretch}{1.3}
\begin{tabular}{lc}
\hline\hline
\textbf{Quantity} & \textbf{Value} \\
\hline
FCI reference energy $E_\text{FCI}$ & $-1.137284$ \\
Nuclear repulsion energy $E_\text{nuc}$ & $0.715104$ \\
Identity coefficient $c_0$ & $-0.812171$ \\
Shifted energy $E'$ & $-1.040218$ \\
Rescaled phase $\phi$ & $339.058$ \\
Recovered energy $E_\text{total}$ & $-1.137105$ \\
Absolute error $|\Delta E|$ & $1.786 \times 10^{-4}$ \\
\hline\hline
\end{tabular}
\end{table}

The recovered energy differs from the FCI reference by
$1.786 \times 10^{-4}$~Hartree, well within the chemical accuracy
threshold of $1.59 \times 10^{-3}$~Hartree~\cite{mcardle2020quantum,
helgaker2000molecular}. In every H$_2$ configuration at $p = 1$, both in
simulation and on real hardware, the most probable measured bitstring is
$k^* = \texttt{00101010011}$, so all of them recover this same energy.
Since the recovered energy depends only on $k^*$, this error is the
discretization error of the $11$-qubit phase register for the chosen
$\tau$, $2\pi|\phi - k^*|/(2^{n_a}\tau)$; the role of the VQC surrogate
is to reproduce the correct dominant bitstring, which it does in all
cases.

\subsection{Key Findings of the Selection Stages}
\label{subsec:results_summary}

The selection stages yield the following findings: (i) the linear
entangler substantially outperforms the full entangler on real
hardware; (ii) $p=1$ is the best number of layers under hardware noise;
(iii) for the linear-entangler circuits, DD degrades hardware
performance at every tested depth, with a penalty that grows with
depth; (iv) the two ans\"atze achieve comparable
hardware performance, and the $R_Y$-$CZ$ ansatz is preferred for its
lower parameter and gate count; and (v) ideal-simulator-trained
parameters generalize better to real hardware than
noisy-simulator-trained parameters. The side-lobe structure of the
hardware distributions is analysed in Section~\ref{sec:sidelobes}, and the behavior of
DD at larger depths ($p \in \{8, 64\}$) in Section~\ref{sec:noise_analysis} of the
Supplementary Material.

\subsection{Validation on Further Molecules}
\label{subsec:results_molecules}

To assess whether the analytical training pipeline and the selected
configuration generalize beyond H$_2$, we applied them without further
modification to LiH, H$_2$O, BeH$_2$, NH$_3$, CH$_4$, O$_2$ and N$_2$.
For each molecule, the Hamiltonian was constructed and mapped as
described in Section~\ref{subsec:mol_system}, the FCI reference energy
was obtained with PySCF~\cite{sun2018pyscf}, and the Dirichlet target
was computed via Eq.~\eqref{eq:dirichlet_explicit} using the
molecule-specific evolution time $\tau$ listed in Table~A1 of the
Supplementary Material. A single-layer linear-entangler $R_Y$-$CZ$
ansatz was trained on the ideal simulator and executed both on the
ideal simulator (Ideal VQC) and on \texttt{ibm\_boston} (Real VQC),
without DD.

\begin{figure}[htbp]
    \centering
    \includegraphics[width=\linewidth]{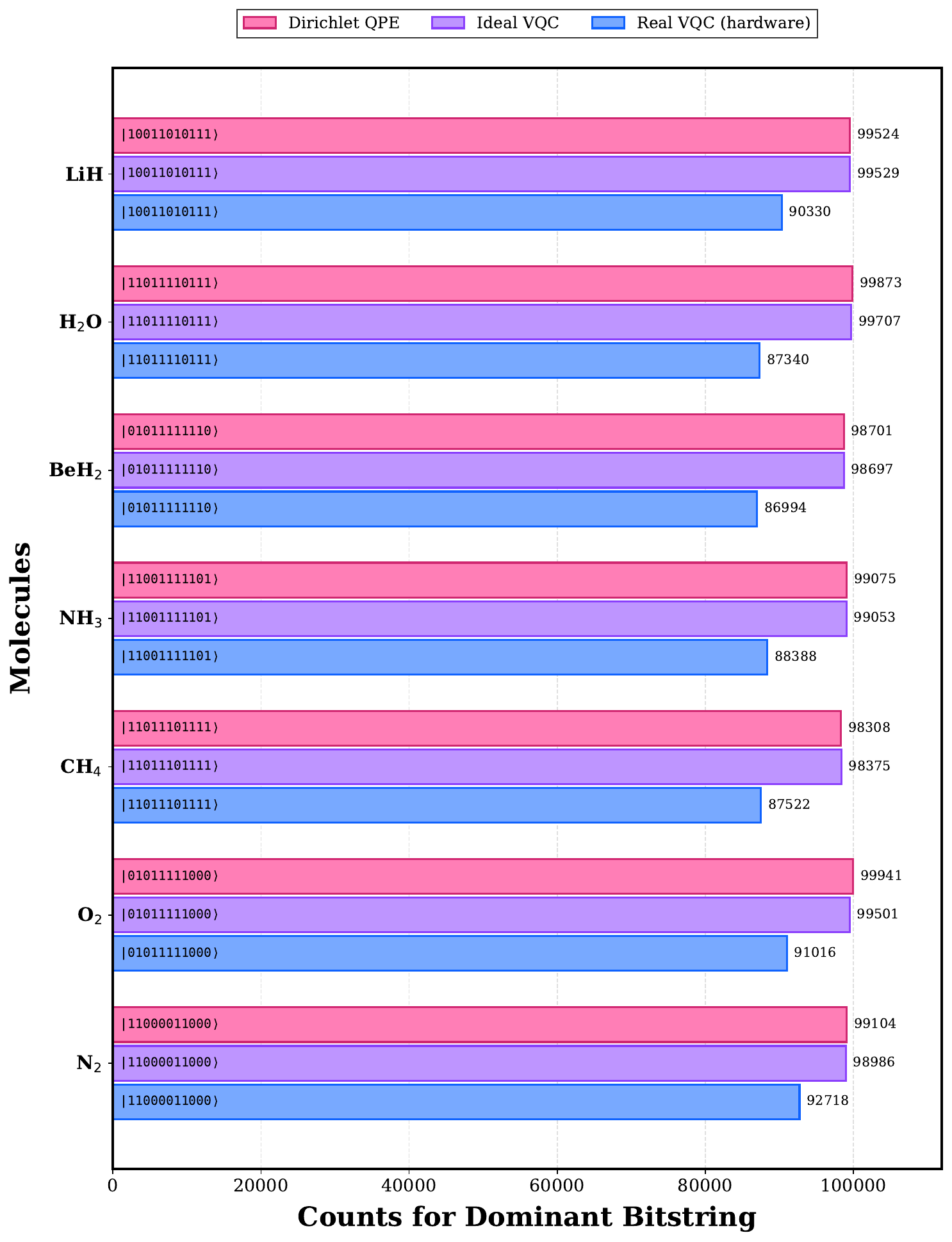}
    \caption{Counts of the dominant bitstring for the analytical
    Dirichlet reference, the ideal-simulator VQC and the hardware VQC
    (\texttt{ibm\_boston}) for seven molecules
    ($N_\text{shots} = 100{,}000$), using the single-layer
    linear-entangler $R_Y$-$CZ$ ansatz with ideal-simulator-trained
    parameters. The dominant bitstring of each molecule is labeled on
    its bars.}
    \label{fig:dominant_peak_molecules}
\end{figure}

Figure~\ref{fig:dominant_peak_molecules} shows, for each molecule, the
count of the dominant bitstring for the analytical Dirichlet reference,
the ideal-simulator VQC and the hardware VQC. For every molecule, the
same dominant bitstring is obtained in all three cases, showing that the
surrogate reproduces the dominant QPE outcome both in simulation and on
real hardware, although hardware execution reduces its count by roughly
$6$--$13\%$ relative to the Dirichlet reference.

\begin{figure}[htbp]
    \centering
    \includegraphics[width=\linewidth]{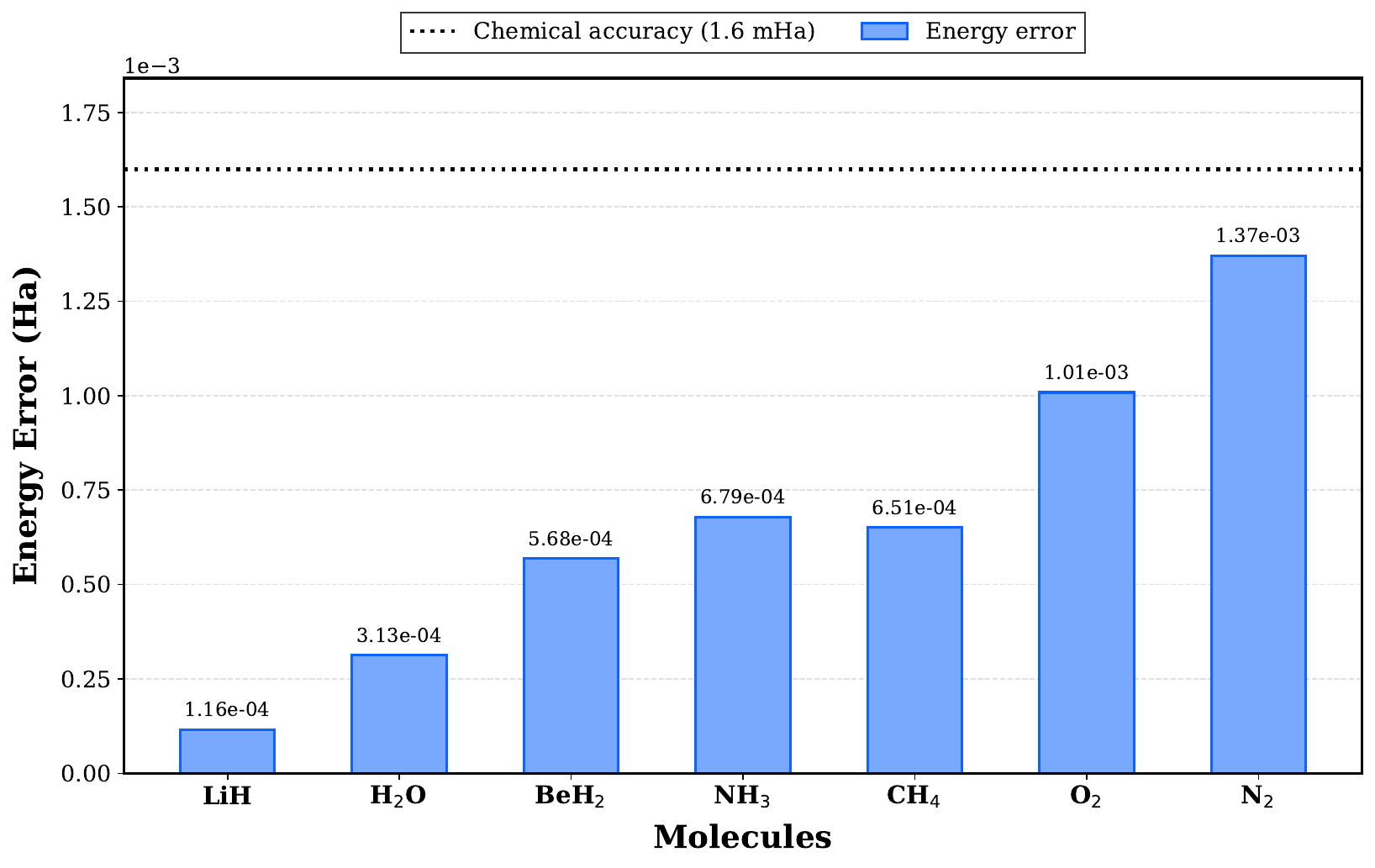}
    \caption{Absolute ground-state energy error $|\Delta E|$ recovered
    from the dominant bitstring for seven molecules. The ideal-simulator
    and hardware VQCs yield the same dominant bitstring and therefore
    the same error. The dotted line marks the chemical accuracy
    threshold.}
    \label{fig:mols_energy_error}
\end{figure}

Figure~\ref{fig:mols_energy_error} reports the absolute energy error
recovered from this dominant bitstring. Because the ideal-simulator and
hardware executions share the same dominant bitstring, the error is
identical for both. All seven molecules achieve energy errors below the
chemical accuracy threshold, ranging from $1.16 \times 10^{-4}$~Hartree
(LiH) to $1.37 \times 10^{-3}$~Hartree (N$_2$). For all eight molecules, including H$_2$, the energy error equals the discretization error of the phase register for the chosen $\tau$. The values of $\tau$ were selected using the FCI reference
energy such that $\phi$ lies close to an integer (see Section~\ref{subsec:molecules} of
the Supplementary Material), so the results demonstrate that the
surrogate reproduces the dominant QPE outcome on hardware for a suitably
chosen evolution time.

\FloatBarrier
\section{Conclusion}
\label{sec:conclusion}

We have presented a variational surrogate framework for the measurement distribution of QPE
in molecular energy estimation, in which the training target is computed analytically from
the Dirichlet kernel using only a reference energy, the number of ancilla qubits and the
evolution time. This reformulates the approach of Liu \textit{et al.}~\cite{liu2024learnqpe}
by removing the need to simulate the QPE circuit, and in particular its controlled-unitary
and inverse-QFT operations, so that the cost of constructing the target shifts from
simulating a QPE circuit to obtaining the reference energy. Through a stepwise selection on
H$_2$ evaluated on IBM Quantum hardware, we found that the linear entangler substantially
outperforms the full entangler, that a single layer performs best under hardware noise, that
the $R_Y$-$CZ$ ansatz matches the $R_Y$-$R_Z$-$CZ$ ansatz with half the parameters, and that
ideal-simulator-trained parameters transfer to hardware better than noisy-simulator-trained
ones. The resulting single-layer linear-entangler $R_Y$-$CZ$ circuit recovered the dominant
QPE bitstring in simulation and on hardware for H$_2$, LiH, H$_2$O, BeH$_2$, NH$_3$, CH$_4$,
O$_2$ and N$_2$, giving energy errors between $1.2 \times 10^{-4}$ and
$1.4 \times 10^{-3}$~Hartree, within chemical accuracy for the chosen evolution times. Beyond
the dominant peak, the side-lobe analysis (Section~\ref{sec:sidelobes} of the Supplementary Material) showed
that the hardware distributions reproduce the near-neighbor decay of the Dirichlet target
with a broadened envelope, together with isolated hardware-induced side lobes absent from the
target. The noise analysis (Section~\ref{sec:noise_analysis} in Supplementary Material) showed that XpXm dynamical decoupling degraded the
linear-entangler circuits at every tested depth, indicating that, for such densely scheduled
circuits, circuit design is a more effective lever for noise resilience than pulse-level
error suppression.

The framework is intended for settings in which QPE enters a larger workflow as a subroutine
and only its measured outcome is passed on. There, the trained VQC reproduces the output
distribution of the deep QPE block with a shallow circuit that can be executed on current
hardware, so that the remaining stages of the workflow can be run and tested before
fault-tolerant QPE is available. The approach has, however, several limitations. First, the
surrogate reproduces only the measurement probabilities of the phase register, not its
complex amplitudes, its correlations with the system register or the post-measurement state
of the system register. It therefore cannot replace QPE where it is used for projective
ground-state preparation~\cite{abrams1999quantum, ge2019faster, lin2020near} or as a coherent
subroutine within a larger quantum algorithm~\cite{kassal2009quantum, temme2011quantum}, as
in the HHL algorithm, in which the phase register is uncomputed rather than
measured~\cite{harrow2009quantum}. Second, training requires the reference energy in advance,
so the surrogate does not by itself determine an unknown energy and is not suited to
standalone energy estimation. Third, the reference energy used here is obtained from FCI,
which becomes classically intractable for larger molecules and basis sets. Fourth, the
recovered energy error equals the discretization error of the phase register, fixed by $n_a$
and $\tau$, so the reported chemical accuracy depends on the choice of $\tau$ and indicates only
that the hardware preserves the dominant bitstring.

These limitations point to the main directions for future work. The FCI reference can be
replaced with scalable approximate methods, such as coupled cluster, and the propagation of
their error into the surrogate can be quantified. A systematic procedure for selecting $\tau$
is needed, for example one based on a lower-accuracy energy estimate or on adaptive
refinement. The surrogate should be tested on hardware at evolution times for which $\phi$
lies far from an integer, where the entangling layer of the VQC becomes essential, and the
trainability of the linear-entangler VQC and its resilience to barren
plateaus~\cite{mcclean2018barren, cerezo2021cost} should be examined for larger ancilla
registers. Finally, the framework can be combined with error-mitigation techniques such as
zero-noise extrapolation~\cite{temme2017error, giurgica2020digital} and readout error
mitigation~\cite{bravyi2021mitigating}, and the analytical target can be extended to multiple
eigenphases for excited-state applications.

\section*{Data Availability}
The data that support the findings of this study are available from the corresponding author upon reasonable request.

\section*{Acknowledgements}
We acknowledge the use of IBM Quantum Credits via the IBM Quantum Startups Program for this work. The views expressed are those of the authors and do not reflect the official policy or position of IBM or the IBM Quantum Platform team. The authors would like to express their appreciation to the advisors at Qclairvoyance Quantum Labs for their support, constructive discussions, and inspiration throughout the preparation of this work. 

\section*{Funding}
No funding was received for this research.

\section*{Conflict of Interest}
The authors declare no competing interests.

\bibliography{references}

% \section*{Supplementary information (optional)}
% If your article requires supplementary information, please include these files for peer-review. Please note that supplementary information will not be edited.

\newpage
\appendix
\section*{Supplementary Material}

% \input{sections/theoretical_background}
% \begin{center}
%     {\Large\bfseries Supplementary Material}
% \end{center}

\setcounter{section}{0}
\renewcommand{\thesection}{\Alph{section}}
\counterwithin*{equation}{section}
\counterwithin*{figure}{section}
\counterwithin*{table}{section}
\renewcommand{\theequation}{\thesection\arabic{equation}}
\renewcommand{\thefigure}{\thesection\arabic{figure}}
\renewcommand{\thetable}{\thesection\arabic{table}}

% =====================================================================
\section{Theoretical Background}
\label{sec:theory}
% =====================================================================

\subsection{Historical Development of Quantum Phase Estimation}
\label{subsec:qpe_history}

The problem of estimating the eigenphase of a unitary operator has a rich
history in quantum algorithm design, with successive refinements progressively
clarifying both the theoretical foundations and the practical circuit
requirements. The earliest formulation is due to Kitaev~\cite{kitaev1995quantum},
who introduced a phase estimation procedure based on repeated single-qubit
measurements and classical post-processing. In Kitaev's original scheme, the
phase $\theta$ is estimated bit by bit using a sequence of
controlled-$U^{2^k}$ operations, each followed by a Hadamard gate and a
single-qubit measurement. While conceptually transparent, this approach
requires many repetitions and does not produce the full probability
distribution over bitstrings in a single circuit execution.

The circuit-based formulation of QPE, which has since become standard, was
introduced and analyzed by Cleve \textit{et al.}~\cite{cleve1998quantum} and
independently described in the textbook treatment by Mosca and
Ekert~\cite{mosca1999hidden}. In this formulation, an ancilla register of
$n_a$ qubits is initialized in the uniform superposition via Hadamard gates,
followed by a sequence of controlled-$U^{2^k}$ operations acting on the
system register initialized in the eigenstate $\ket{\psi}$. An inverse
Quantum Fourier Transform (QFT$^\dagger$) is then applied to the ancilla
register, and measurement in the computational basis yields an $n_a$-bit
approximation to $\theta$ with high
probability~\cite{cleve1998quantum, nielsen2000quantum}. The textbook by
Nielsen and Chuang~\cite{nielsen2000quantum} provided the definitive
pedagogical treatment, establishing the resource requirements and error
analysis that are now widely cited. Specifically, to estimate $\theta$ with
precision $\epsilon$ and success probability at least $1 - \delta$, one
requires
\begin{equation}
    n_a = n_\text{bits} + \left\lceil \log_2\!\left(2 +
    \frac{1}{2\delta}\right) \right\rceil
    \label{eq:qpe_qubits}
\end{equation}
ancilla qubits, where $n_\text{bits}$ is the number of bits of precision
required to resolve the phase to within $\epsilon = 2^{-n_\text{bits}}$.
The QFT$^\dagger$ contributes $\mathcal{O}(n_a^2)$ gates, while the
controlled-$U^{2^j}$ operations, whose cost grows exponentially with
$n_a$, dominate the total depth for molecular Hamiltonians
(Section~\ref{subsec:qpe_circuit}), rendering direct hardware execution
impractical on NISQ devices.

The application of QPE to quantum chemistry was pioneered by Aspuru-Guzik
\textit{et al.}~\cite{aspuru2005simulated}, who showed that the ground-state
energy of a molecular Hamiltonian $H$ can be extracted by applying QPE to
the time-evolution operator $U = e^{-iH\tau}$, with the eigenphase encoding
the energy as $\theta = |E|\tau/2\pi$. This sparked a sustained research effort
into quantum algorithms for electronic structure
calculation~\cite{babbush2018encoding, reiher2017elucidating,
wecker2015progress, mcardle2020quantum}, culminating in resource estimates
for fault-tolerant QPE on industrially relevant
molecules~\cite{reiher2017elucidating, babbush2018encoding}. Subsequent
work by Berry \textit{et al.}~\cite{berry2019qubitization} introduced
qubitization and quantum signal processing as more efficient alternatives
to Trotterized time evolution for Hamiltonian simulation within QPE,
further reducing the gate complexity of the controlled-$U$ operations.

More recently, attention has shifted to phase estimation protocols tailored
for the pre-fault-tolerant regime. Lin and Tong~\cite{lin2022heisenberg}
introduced a rejection-sampling-based phase estimation algorithm that
achieves Heisenberg-limited precision scaling with early fault-tolerant
hardware. Parallel developments in robust phase
estimation~\cite{kimmel2015robust} and Bayesian phase
estimation~\cite{wiebe2016efficient} have explored statistical approaches
to extracting phase information from shallow circuits with fewer ancilla
qubits. Liu \textit{et al.}~\cite{liu2024learnqpe} proposed a distinct
direction in which the QPE circuit is replaced entirely by a trained
variational surrogate that reproduces the QPE measurement distribution at
greatly reduced circuit depth. This idea directly motivates the
analytically grounded framework introduced in the present work.

\subsection{Quantum Phase Estimation: Circuit Formalism and Depth Analysis}
\label{subsec:qpe_circuit}

In the standard circuit formulation, QPE estimates the phase
$\theta \in [0,1)$ of a unitary $U$ satisfying
$U\ket{\psi} = e^{2\pi i\theta}\ket{\psi}$, using an $n_a$-qubit ancilla
register and a system register initialized in the eigenstate
$\ket{\psi}$~\cite{cleve1998quantum, nielsen2000quantum}. The algorithm
proceeds as follows:

\begin{enumerate}
    \item Apply Hadamard gates to all $n_a$ ancilla qubits to prepare the
    uniform superposition
    $\frac{1}{\sqrt{2^{n_a}}}\sum_{k=0}^{2^{n_a}-1}\ket{k}$.

    \item Apply the controlled-$U^{2^j}$ gate for
    $j = 0, 1, \ldots, n_a - 1$, where the $j$-th ancilla qubit serves
    as the control.

    \item Apply the inverse Quantum Fourier Transform QFT$^\dagger$ to
    the ancilla register.

    \item Measure the ancilla register in the computational basis.
\end{enumerate}

The resulting measurement outcome $k \in \{0, 1, \ldots, 2^{n_a} - 1\}$
is obtained with probability~\cite{nielsen2000quantum, cleve1998quantum}
\begin{equation}
    P(k;\phi) = \frac{1}{2^{2n_a}}
    \frac{\sin^2\!\left(\pi(\phi - k)\right)}
    {\sin^2\!\left(\dfrac{\pi(\phi - k)}{2^{n_a}}\right)},
    \label{eq:qpe_prob}
\end{equation}
where $\phi = 2^{n_a}\theta$ is the rescaled phase. When $\phi$ is
exactly an integer, the distribution collapses to a Kronecker delta at
$k = \phi$, recovering the phase exactly in a single shot. For
non-integer $\phi$, the distribution takes the form of a squared
Dirichlet kernel, sharply peaked at $k^* = \lfloor\phi\rceil$ (the
nearest integer) with characteristic decaying side
lobes~\cite{cleve1998quantum}.

In the context of quantum chemistry, $U = e^{-iH\tau}$ is the
time-evolution operator under the molecular Hamiltonian $H$, and the
eigenphase encodes the energy eigenvalue~\cite{aspuru2005simulated,
wecker2015progress}. Constant terms in the Hamiltonian contribute only a
global phase and carry no information, so they are removed before the
phase is defined. Writing the qubit-mapped electronic Hamiltonian as
$H_\text{el} = c_0\,I + H'$, where $c_0$ is the coefficient of
the identity term, and the total energy as $E = E_\text{el} +
E_\text{nuc}$, where $E_\text{nuc}$ is the nuclear repulsion energy, the
phase is set by the eigenvalue of the traceless part $H'$,
\begin{equation}
    E' = E - E_\text{nuc} - c_0,
    \qquad
    \theta = \frac{|E'|\,\tau}{2\pi},
    \qquad
    \phi = 2^{n_a}\,\frac{|E'|\,\tau}{2\pi}.
    \label{eq:shifted_energy}
\end{equation}
For the phase to be uniquely defined, the evolution time must satisfy
$|E'|\tau < 2\pi$, so that $\theta \in [0,1)$ and the phase does not
wrap around; the values of $\tau$ used in this work are discussed in
Section~\ref{subsec:molecules}. Since $E'$ is negative for the ground states considered
here, the dominant measurement outcome $k^*$ yields the total-energy
estimate
\begin{equation}
    E_\text{est} = E_\text{nuc} + c_0 - \frac{2\pi k^*}{2^{n_a}\,\tau}.
    \label{eq:energy_est}
\end{equation}
Because $E_\text{nuc}$ and $c_0$ are known exactly, the energy error
$|E_\text{est} - E_\text{ref}|$ is identical for the shifted and the
total energy.

\subsubsection*{QPE Circuit Depth for Molecular Hamiltonians}

The circuit depth of QPE for molecular systems is determined by two
contributions: the inverse QFT and the controlled time-evolution
operators. For a molecule with $n_s$ spin-orbitals, the molecular
Hamiltonian after Jordan-Wigner mapping contains
$\mathcal{O}(n_s^4)$ Pauli terms, arising from the two-electron
integrals~\cite{mcardle2020quantum, helgaker2000molecular}.
Implementing the time-evolution operator $U = e^{-iH\tau}$ via
first-order Trotterization with $m$ Trotter steps incurs a gate
complexity of
\begin{equation}
    D_\text{Trotter} \sim \mathcal{O}(m \cdot n_s^4),
    \label{eq:trotter_depth}
\end{equation}
since each of the $\mathcal{O}(n_s^4)$ Pauli exponentials must be
applied sequentially within each Trotter
step~\cite{wecker2015progress, mcardle2020quantum}. For the $j$-th
controlled-$U^{2^j}$ operation in the QPE circuit, this Trotterized
block is repeated $2^j$ times, giving a per-ancilla depth of
$\mathcal{O}(2^j \cdot m \cdot n_s^4)$. Summing over all $n_a$ ancilla
qubits yields the total QPE circuit depth:
\begin{equation}
    D_\text{QPE} = \sum_{j=0}^{n_a - 1}
    \mathcal{O}\!\left(2^j \cdot m \cdot n_s^4\right)
    = \mathcal{O}\!\left(2^{n_a} \cdot m \cdot n_s^4\right).
    \label{eq:qpe_depth}
\end{equation}
Substituting the ancilla count requirement from
Eq.~\eqref{eq:qpe_qubits}, where $n_a \sim \mathcal{O}(\log(1/\epsilon))$
for precision $\epsilon$, gives $2^{n_a} \sim \mathcal{O}(1/\epsilon)$,
and the total depth becomes
\begin{equation}
    D_\text{QPE} \sim \mathcal{O}\!\left(\frac{m \cdot n_s^4}{\epsilon}
    \right).
    \label{eq:qpe_depth_final}
\end{equation}
This expression reveals the three-way scaling challenge inherent to
molecular QPE: the depth grows quartically with the number of
spin-orbitals $n_s$, linearly with the number of Trotter steps $m$,
and inversely with the required precision $\epsilon$. For molecules
beyond hydrogen such as LiH ($n_s = 12$), BeH$_2$ ($n_s = 14$), or
water ($n_s = 14$), the $\mathcal{O}(n_s^4)$ scaling renders the QPE
circuit depth prohibitively large even for moderate precision
requirements on current
hardware~\cite{wecker2015progress, mcardle2020quantum}.

The same scaling also limits the classical simulation of QPE circuits.
A classical state-vector simulator requires memory scaling as
$\mathcal{O}(2^{n_s + n_a})$ and runtime proportional to
$D_\text{QPE}$, making exact simulation of QPE for anything beyond the
smallest molecules computationally infeasible on classical hardware.
This is the fundamental bottleneck of prior surrogate
approaches~\cite{liu2024learnqpe} that rely on classical QPE simulation
to generate training targets. The analytical Dirichlet kernel framework
introduced in this work circumvents this bottleneck: it requires only a
reference ground-state energy $E_\text{ref}$, the ancilla count $n_a$
and the time evolution parameter $\tau$, together with an
$\mathcal{O}(2^{n_a})$ classical evaluation of the Dirichlet
distribution, which is lightweight compared to full circuit simulation.

\subsection{The Dirichlet Kernel and the Analytical Target Distribution}
\label{subsec:dirichlet}

A foundational contribution of the framework introduced in this work is
the replacement of QPE circuit simulation with a fully analytical
computation of the training target, enabled by the closed-form structure
of the QPE measurement distribution. Given a reference ground-state
energy $E_\text{ref}$ for the chosen basis, together with the number of
ancilla qubits $n_a$ and the time evolution parameter $\tau$, the
rescaled phase follows from Eq.~\eqref{eq:shifted_energy} as
$\phi = 2^{n_a}|E'_\text{ref}|\,\tau/(2\pi)$, with
$E'_\text{ref} = E_\text{ref} - E_\text{nuc} - c_0$.
In this work, $E_\text{ref}$ is obtained from a Full Configuration
Interaction (FCI) calculation with PySCF~\cite{sun2018pyscf}, and
$E_\text{nuc}$ and $c_0$ are taken from the same molecular
Hamiltonian that is mapped to qubits. Assuming the system register is prepared
in the exact ground state, the QPE measurement distribution is then
given exactly by the squared Dirichlet kernel of
Eq.~\eqref{eq:qpe_prob},
\begin{equation}
    P_\text{gt}(k;\phi) = \frac{1}{2^{2n_a}}
    \frac{\sin^2\!\left(\pi(\phi - k)\right)}
    {\sin^2\!\left(\dfrac{\pi(\phi - k)}{2^{n_a}}\right)},
    \quad k = 0,1,\ldots,2^{n_a}-1.
    \label{eq:dirichlet}
\end{equation}
This expression is normalized,
$\sum_{k=0}^{2^{n_a}-1}P_\text{gt}(k;\phi) = 1$, and is sharply peaked
at $k^* = \lfloor \phi \rceil$~\cite{cleve1998quantum}. The probability
carried by the dominant bitstring depends on how close $\phi$ is to an
integer: it equals unity when $\phi$ is an integer and is bounded below
by $4/\pi^2 \approx 0.405$ when $\phi$ lies midway between two
integers~\cite{nielsen2000quantum}. For the H$_2$ configuration with
$n_a = 11$ ancilla qubits studied here, the dominant bitstring carries
$98.9\%$ of the probability mass. The remaining
probability is distributed across neighboring side lobes, whose accurate
characterization requires adequate sampling depth. In all experiments
reported in this work, we use $N_\text{shots} = 100{,}000$ measurement
shots uniformly across every experimental configuration, for both
simulation and hardware, ensuring statistically consistent and directly
comparable histogram estimates between the VQC output distribution and
the analytical Dirichlet target across all stages of the investigation.

The key computational advantage is that $P_\text{gt}(k;\phi)$ can be
evaluated entirely classically for all $2^{n_a}$ bitstrings in
$\mathcal{O}(2^{n_a})$ time, requiring only $E_\text{ref}$, $n_a$ and
$\tau$ as inputs, with no quantum circuit execution or simulation
required to generate the target. This stands in sharp contrast to prior
surrogate approaches~\cite{liu2024learnqpe}, in which the training
target must be generated by running the full QPE circuit on a classical
state-vector simulator. As established in
Section~\ref{subsec:qpe_circuit}, the memory and runtime cost of such
simulation scales as $\mathcal{O}(2^{n_s + n_a})$ and
$\mathcal{O}(2^{n_a} \cdot m \cdot n_s^4)$ respectively, becoming
infeasible for any molecule beyond the smallest benchmark systems. In
contrast, once $E_\text{ref}$ is available, the cost of the Dirichlet
target is independent of the molecular size $n_s$, and since the VQC
acts only on the $n_a$ ancilla qubits, the cost of training the VQC is
likewise independent of $n_s$. The only molecule-dependent classical
cost is that of obtaining $E_\text{ref}$: FCI, used here, scales
exponentially with system size, but the framework accepts a reference
energy from any sufficiently accurate classical method.

\subsection{Variational Quantum Circuits}
\label{subsec:vqc}

Variational Quantum Circuits (VQCs), also known as parameterized quantum
circuits (PQCs), are quantum circuits $U(\boldsymbol{\theta})$ whose gate
parameters $\boldsymbol{\theta} = \{\theta_1,\ldots,\theta_M\}$ are
optimized by a classical outer loop to minimize a scalar cost
function~\cite{cerezo2021variational, bharti2022noisy}. The theoretical
foundations of VQCs draw on the expressibility of parameterized
unitaries~\cite{sim2019expressibility} and the trainability landscape of
quantum cost functions~\cite{mcclean2018barren, cerezo2021cost}. VQCs
have been applied to variational
eigensolvers~\cite{peruzzo2014variational, kandala2017hardware}, quantum
classifiers~\cite{biamonte2017quantum}, quantum generative
models~\cite{benedetti2019generative}, and quantum reinforcement
learning~\cite{chen2020variational}, among many other
tasks~\cite{cerezo2021variational}.

The architecture of a VQC consists of alternating layers of single-qubit
rotation gates and two-qubit entangling gates. Two entangler topologies
are standard: the \textit{linear} entangler, coupling each qubit to its
nearest neighbor and yielding depth $\mathcal{O}(p \cdot n_a)$ for $p$
layers and $n_a$ ancilla qubits; and the \textit{full} entangler,
coupling all pairs and yielding depth
$\mathcal{O}(p \cdot n_a^2)$~\cite{liu2024learnqpe,
sim2019expressibility}. Under ideal (noiseless) conditions, the full
entangler achieves greater expressibility for a fixed layer count;
however, under realistic noise, its deeper circuits accumulate
substantially more error, and the linear entangler is the preferred
choice for NISQ hardware~\cite{liu2024learnqpe, cerezo2021variational}.
In the present framework, the VQC serves as a linearly scaling surrogate
for the QPE circuit with a depth of $\mathcal{O}(p \cdot n_a)$, compared
to $\mathcal{O}(2^{n_a} \cdot m \cdot n_s^4)$ for the full QPE circuit,
and is trained using the analytical Dirichlet target before deployment
on real quantum hardware.

We investigate two ansatz designs. The \textbf{$R_Y$-$R_Z$-$CZ$ ansatz}
applies $R_Y$ and $R_Z$ rotations to every qubit in each layer, followed
by $CZ$ entangling gates according to the chosen topology, giving
$2(p+1)n_a$ variational parameters. The \textbf{$R_Y$-$CZ$ ansatz}
applies only $R_Y$ rotations per layer, reducing the parameter count to
$(p+1)n_a$ and further decreasing circuit depth while retaining the
entanglement structure necessary for learning the QPE distribution. Both
ans\"atze are initialized with all variational parameters set to zero.

\subsection{Cost Function and Classical Optimization}
\label{subsec:cost}

The VQC parameters are optimized by minimizing the cost function
\begin{equation}
    \mathcal{L}(\boldsymbol{\theta}) = \sum_{k=0}^{2^{n_a}-1}
    \sqrt{\bigl[P_{\boldsymbol{\theta}}(k) -
    P_\text{gt}(k;\phi)\bigr]^2},
    \label{eq:cost}
\end{equation}
which is the $L^1$ norm of elementwise absolute differences between the
VQC output distribution $P_{\boldsymbol{\theta}}(k)$ and the analytical
Dirichlet target $P_\text{gt}(k;\phi)$. Since the training target is
computed analytically rather than by circuit execution, the optimization
loop is entirely decoupled from any QPE simulation overhead, and the
only quantum resource consumed during training is the VQC circuit
evaluation itself. Optimization is performed using the Constrained
Optimization BY Linear Approximation (COBYLA)
algorithm~\cite{powell1994direct}, a gradient-free method that constructs
a linear approximation to the objective function within a trust region
and updates parameters accordingly. A critical practical advantage of
COBYLA is that it requires only a single circuit execution per
optimization iteration, avoiding the $\mathcal{O}(M)$ circuit evaluations
per step required by gradient-based methods such as the parameter-shift
rule~\cite{mitarai2018quantum, schuld2019evaluating}, where $M$ is the
number of variational parameters in $\boldsymbol{\theta}$ (e.g.,
$M = 2(p+1)n_a$ for the $R_Y$-$R_Z$-$CZ$ ansatz and $M = (p+1)n_a$ for the
$R_Y$-$CZ$ ansatz). This makes COBYLA
particularly well-suited to VQC training on NISQ devices where circuit
execution time is a primary
bottleneck~\cite{liu2024learnqpe, cerezo2021variational}. At each COBYLA
iteration, the VQC output distribution $P_{\boldsymbol{\theta}}(k)$ is
estimated using $N_\text{shots} = 100{,}000$ measurement shots, yielding
normalized probability estimates
\begin{equation}
    P_{\boldsymbol{\theta}}(k) = \frac{c_k}{N_\text{shots}},
    \quad k = 0, 1, \ldots, 2^{n_a} - 1,
    \label{eq:prob_estimate}
\end{equation}
where $c_k$ denotes the count of bitstring $k$ in the measurement record.
The training target in Eq.~\eqref{eq:cost} is the exact probability
vector $P_\text{gt}(k;\phi)$ of Eq.~\eqref{eq:dirichlet}, not a sampled
histogram. For visual comparison with the measured histograms and for
the side-lobe analysis of Section~\ref{sec:sidelobes}, the Dirichlet
target is additionally expressed on the same $N_\text{shots} = 100{,}000$ scale by truncating
the expected count $N_\text{shots}\,P_{\text{gt}}(k;\phi)$ of each
bitstring to its integer part. Bitstrings whose expected count is
below one therefore contribute zero, so the resulting reference histogram contains
$99{,}948$ counts, of which $98{,}890$ fall on the dominant bitstring
\texttt{00101010011}. For bin
probabilities of order $q \sim 10^{-3}$, this shot count yields an
expected sample count of $N_\text{shots} \cdot q \approx 100$, sufficient
to suppress large relative statistical fluctuations
($\sigma_q \approx \sqrt{q(1-q)/N_\text{shots}} \approx 1.0 \times
10^{-4}$) and to ensure stable estimation of divergence-based metrics
such as the Jensen--Shannon divergence (JSD), which are sensitive to tail
undersampling.

\subsection{Distributional Distance Metrics}
\label{subsec:metrics}

To rigorously quantify the agreement between the learned VQC distribution
$P \equiv P_{\boldsymbol{\theta}}(k)$ and the analytical Dirichlet target
$Q \equiv P_\text{gt}(k;\phi)$, we employ four complementary
distributional distance measures. Together, these metrics probe different
geometric and information-theoretic aspects of the distributional
discrepancy, providing a more comprehensive evaluation of VQC performance
than the single $\chi^2$ statistic employed in prior
work~\cite{liu2024learnqpe}.

\subsubsection*{Hellinger Distance}

The Hellinger distance~\cite{hellinger1909neue} is a symmetric, bounded
measure of dissimilarity between two probability distributions, defined
as
\begin{equation}
    H(P,Q) = \frac{1}{\sqrt{2}}\left\|\sqrt{P} - \sqrt{Q}\right\|_2
    = \sqrt{1 - \sum_k \sqrt{P(k)\,Q(k)}}\,.
    \label{eq:hellinger}
\end{equation}
It satisfies $H(P,Q) \in [0,1]$, with $H = 0$ if and only if $P = Q$,
and $H = 1$ if and only if $P$ and $Q$ have mutually disjoint support.
The Hellinger distance is related to the Bhattacharyya coefficient
$BC(P,Q) = \sum_k\sqrt{P(k)Q(k)}$ by
$H^2 = 1 - BC$~\cite{bhattacharyya1943measure}, and is metrically
equivalent to the total variation distance up to constant factors. Its
square, $H^2$, is directly related to the fidelity between quantum states
when $P$ and $Q$ are treated as classical probability vectors, making it
a natural choice for benchmarking quantum
distributions~\cite{nielsen2000quantum}. A key practical advantage of the
Hellinger distance is its robustness to zero-probability bins, which
makes it well-suited for comparing finite-shot probability distributions
that may contain outcomes with zero observed counts.

\subsubsection*{Fidelity Error}

The classical fidelity between two probability distributions $P$ and $Q$
is defined as
\begin{equation}
    F(P,Q) = \left(\sum_k \sqrt{P(k)\,Q(k)}\right)^2 = BC(P,Q)^2\,,
    \label{eq:fidelity}
\end{equation}
in analogy with the quantum state fidelity
$F(\rho,\sigma) = \left(\mathrm{Tr}\sqrt{\sqrt{\rho}\sigma\sqrt{\rho}}
\right)^2$~\cite{nielsen2000quantum, jozsa1994fidelity}. For diagonal
density matrices, whose eigenvalues form classical probability vectors,
the quantum fidelity reduces to
Eq.~\eqref{eq:fidelity}~\cite{nielsen2000quantum}. The fidelity satisfies
$F \in [0,1]$, with $F = 1$ indicating identical distributions and
$F = 0$ indicating disjoint support. The fidelity error is defined as
\begin{equation}
    \epsilon_F = 1 - F(P,Q) = 1 - \left(\sum_k
    \sqrt{P(k)\,Q(k)}\right)^2\,,
    \label{eq:fidelity_error}
\end{equation}
and measures the departure of the learned distribution from perfect
agreement with the target. It is directly related to the Hellinger
distance by $\epsilon_F = 2H^2 - H^4$, so the two metrics carry
complementary but overlapping information. The fidelity error is
particularly informative in the near-perfect regime ($H \ll 1$), where
$\epsilon_F \approx 2H^2$ amplifies small differences that would be
less visible in the Hellinger distance alone.

\subsubsection*{Total Variation Distance}

The total variation distance (TVD) is the $L^1$-based metric on
probability distributions~\cite{levin2017markov}:
\begin{equation}
    \mathrm{TVD}(P,Q) = \frac{1}{2}\sum_k |P(k) - Q(k)|.
    \label{eq:tvd}
\end{equation}
It has the operational interpretation as the maximum possible difference
in probability that $P$ and $Q$ can assign to any event:
$\mathrm{TVD}(P,Q) = \sup_S |P(S) - Q(S)|$~\cite{levin2017markov}. It
also determines the optimal probability of distinguishing $P$ from $Q$
given a single sample, $\frac{1}{2}\left(1 + \mathrm{TVD}(P,Q)\right)$,
providing a direct operational meaning in the context of hypothesis
testing. The TVD satisfies $\mathrm{TVD}(P,Q) \in [0,1]$ and is related
to the Hellinger distance by
$H^2 \leq \mathrm{TVD} \leq \sqrt{2}\,H$~\cite{tsybakov2009introduction}.
In the present setting, the TVD is particularly sensitive to discrepancies
in the dominant peak of the QPE distribution, since the dominant
bitstring contributes a disproportionately large share of the $L^1$ norm
when the distributions are highly localized.

\subsubsection*{Jensen--Shannon Divergence}

The Jensen--Shannon divergence (JSD) is a symmetrized and smoothed version
of the Kullback--Leibler (KL)
divergence~\cite{kullback1951information, endres2003new}, defined as
\begin{equation}
    \mathrm{JSD}(P\|Q) = \frac{1}{2}D_\mathrm{KL}(P\|M) +
    \frac{1}{2}D_\mathrm{KL}(Q\|M),
    \quad M = \frac{P+Q}{2},
    \label{eq:jsd}
\end{equation}
where $D_\mathrm{KL}(P\|M) = \sum_k P(k)\ln\frac{P(k)}{M(k)}$ is the
KL divergence. Unlike the KL divergence, the JSD is symmetric, always
finite, and bounded: $\mathrm{JSD}(P\|Q) \in [0, \ln 2]$ for the
natural-logarithm convention~\cite{endres2003new}. Endres and
Schindelin~\cite{endres2003new} proved that the square root of the JSD
is a proper metric on the space of probability distributions, satisfying
the triangle inequality. In our analysis, we report
$\sqrt{\mathrm{JSD}/\ln 2}$, which normalizes the metric to $[0,1]$ and
facilitates direct comparison with the Hellinger distance and TVD. The
JSD is particularly sensitive to discrepancies in the low-probability
tail regions of the distributions, where small absolute differences
produce large relative changes in the logarithmic terms. This makes the
JSD a stringent metric for evaluating whether the VQC accurately
captures not only the dominant QPE peak but also the surrounding
side-lobe structure.

\subsection{Molecular Hamiltonian and Qubit Mapping}
\label{subsec:hamiltonian}

The electronic structure of a molecule is governed by the molecular
Hamiltonian, which in second quantization takes the
form~\cite{mcardle2020quantum, helgaker2000molecular, anurag2026}
\begin{equation}
    \hat{H} = \sum_{pq} h_{pq}\,\hat{a}_p^\dagger \hat{a}_q
    + \frac{1}{2}\sum_{pqrs} h_{pqrs}\,
    \hat{a}_p^\dagger \hat{a}_q^\dagger \hat{a}_r \hat{a}_s
    + E_\text{nuc},
    \label{eq:second_quant}
\end{equation}
where $\hat{a}_p^\dagger$ and $\hat{a}_p$ are fermionic creation and
annihilation operators, $h_{pq}$ and $h_{pqrs}$ are one- and two-electron
integrals computed from the chosen molecular orbital basis spanning $n_s$
spin-orbitals, and $E_\text{nuc}$ is the nuclear repulsion
energy~\cite{mcardle2020quantum, anurag2026}. For quantum hardware
implementation, the fermionic operators must be mapped to qubit
operators. The Jordan--Wigner (JW)
transformation~\cite{jordan1928paulische, anurag2026} achieves this via
\begin{equation}
    \hat{a}_p^\dagger \rightarrow
    \frac{1}{2}(\hat{X}_p - i\hat{Y}_p)
    \bigotimes_{q<p}\hat{Z}_q,
    \qquad
    \hat{a}_p \rightarrow
    \frac{1}{2}(\hat{X}_p + i\hat{Y}_p)
    \bigotimes_{q<p}\hat{Z}_q,
    \label{eq:jw}
\end{equation}
where $\hat{X}_p, \hat{Y}_p, \hat{Z}_p$ denote Pauli operators acting on
qubit $p$. Under the JW mapping, the molecular Hamiltonian with $n_s$
spin-orbitals maps to $n_s$ qubits and generates $\mathcal{O}(n_s^4)$
Pauli terms, directly giving rise to the circuit depth scaling derived in
Section~\ref{subsec:qpe_circuit}. Alternative mappings such as the
Bravyi--Kitaev (BK) transformation~\cite{seeley2012bravyi, anurag2026}
offer improved locality at the cost of additional encoding complexity.

In this work, the second-quantized Hamiltonian of each molecule is
constructed with \texttt{Qiskit Nature}~\cite{qiskit2023} from PySCF
integrals and mapped to qubits with the Jordan--Wigner transformation,
without any further qubit reduction; for H$_2$ in the STO-3G basis, for
example, the mapped Hamiltonian acts on $n_s = 4$ qubits. Because the
QPE circuit is never constructed or simulated in the present framework,
the mapped Hamiltonian enters only through the coefficient $c_0$ of its
identity term, which, together with the FCI energy and the nuclear
repulsion energy, defines the shifted energy $E'$ of
Eq.~\eqref{eq:shifted_energy}. Qubit-reduction techniques such as
$\mathbb{Z}_2$ symmetry tapering~\cite{bravyi2017tapering}, which would
reduce the cost of the Hamiltonian-simulation block in a QPE circuit,
are therefore not required. Note that such techniques modify the
identity coefficient, so the value of $c_0$, and hence of $\phi$,
depends on the qubit Hamiltonian from which it is taken; all values
reported here use the untapered Jordan--Wigner Hamiltonian.

\subsection{Molecular Systems and Simulation Parameters}
\label{subsec:molecules}

Table~\ref{tab:molecule_params} lists the molecules studied in this
work, together with the number of spin-orbitals $n_s$ in the STO-3G
basis, the spin multiplicity parameter $2S$, the evolution time $\tau$
used to construct the Dirichlet target of Eq.~\eqref{eq:dirichlet}, the resulting dominant bitstring $k^*$,
and the energy error. All calculations use the STO-3G basis and a
neutral charge. All molecules are treated as closed-shell singlets
($2S = 0$), except O$_2$, which is treated in its triplet ground state
($2S = 2$). The number of ancilla qubits is $n_a = 11$ throughout.

The evolution time $\tau$ was selected empirically for each molecule,
using the FCI reference energy. All selected values satisfy
$|E'_\text{ref}|\,\tau < 2\pi$, so the encoded phase lies in $[0,1)$
without wrapping, but this condition alone does not determine $\tau$ as for a fixed $n_a$, different admissible values of $\tau$ place the
rescaled phase $\phi$ at different distances from the nearest integer
and therefore yield different discretization errors (see below). The
values in Table~\ref{tab:molecule_params} were chosen such that $\phi$
lies close to an integer, which keeps the resulting energy error within
chemical accuracy. Because this choice uses the reference energy, the
reported errors should be read as those achievable for a suitably
chosen $\tau$; a systematic procedure for selecting $\tau$ without
prior knowledge of the energy is left for future work.

The energy error is obtained by converting the most probable bitstring
of the measured VQC distribution into an energy via
Eq.~\eqref{eq:energy_est} and comparing it with the FCI reference,
$|\Delta E| = |E_\text{est} - E_\text{ref}|$. For every molecule, the
ideal-simulator and hardware executions of the VQC yield the same most
probable bitstring, which coincides with the dominant bitstring $k^*$ of
the Dirichlet target, so the reported error applies to both. Because
$E_\text{est}$ is determined by $k^*$ alone, this error equals the
discretization error of the $n_a$-qubit phase register,
$|\Delta E| = 2\pi|\phi - k^*|/(2^{n_a}\tau)$, and is therefore set by
$n_a$, $\tau$ and the position of $\phi$ relative to the nearest
integer.

The Cartesian coordinates of all molecules are given in
Table~\ref{tab:molecule_geometries}. For H$_2$, a bond length of
$0.74$~\AA\ is used; the geometries of the remaining molecules are the
CCSD/STO-3G-optimized structures reported in Ref.~\cite{anurag2026}.

\begin{table}[htbp]
\centering
\caption{Molecules studied, number of spin-orbitals $n_s$ in the STO-3G basis (equal to the
number of system qubits under the Jordan-Wigner mapping), spin parameter $2S$, evolution time
$\tau$, dominant bitstring $k^*$, which is common to the Dirichlet target and to the
ideal-simulator and hardware VQC distributions, and energy error $|\Delta E|$ with respect to
FCI. All molecules are neutral, with $n_a = 11$ ancilla qubits. The chemical accuracy
threshold is $1.59\times10^{-3}$~Ha.}
\label{tab:molecule_params}
\renewcommand{\arraystretch}{1.25}
\begin{tabular}{lccccc}
\hline\hline
\textbf{Molecule} & $\boldsymbol{n_s}$ & $\boldsymbol{2S}$ &
$\boldsymbol{\tau}$ \textbf{(a.u.)} & $\boldsymbol{k^{*}}$ &
$\boldsymbol{|\Delta E|}$ \textbf{(Ha)} \\
\hline
H$_2$   & 4  & 0 & 1.0      & \texttt{00101010011} & $1.786\times10^{-4}$ \\
LiH     & 12 & 0 & 1.01     & \texttt{10011010111} & $1.16\times10^{-4}$ \\
H$_2$O  & 14 & 0 & 0.1929   & \texttt{11011110111} & $3.13\times10^{-4}$ \\
BeH$_2$ & 14 & 0 & 0.34     & \texttt{01011111110} & $5.68\times10^{-4}$ \\
NH$_3$  & 16 & 0 & 0.24     & \texttt{11001111101} & $6.79\times10^{-4}$ \\
CH$_4$  & 18 & 0 & 0.339    & \texttt{11011101111} & $6.51\times10^{-4}$ \\
O$_2$   & 20 & 2 & 0.040401 & \texttt{01011111000} & $1.01\times10^{-3}$ \\
N$_2$   & 20 & 0 & 0.116999 & \texttt{11000011000} & $1.37\times10^{-3}$ \\
\hline\hline
\end{tabular}
\end{table}

\begin{table}[htbp]
\centering
\caption{Cartesian coordinates (\AA) of the molecular geometries used in
this work. The H$_2$ geometry corresponds to a bond length of
$0.74$~\AA; all other geometries are CCSD/STO-3G-optimized structures
from Ref.~\cite{anurag2026}. Only interatomic distances and angles
affect the energies; the absolute positions are those of the original
source.}
\label{tab:molecule_geometries}
\small
\renewcommand{\arraystretch}{1.1}
\begin{tabular}{llrrr}
\hline\hline
\textbf{Molecule} & \textbf{Atom} & $\boldsymbol{x}$ & $\boldsymbol{y}$ & $\boldsymbol{z}$ \\
\hline
H$_2$ & H & $0.0000$ & $0.0000$ & $0.0000$ \\
 & H & $0.0000$ & $0.0000$ & $0.7400$ \\
\hline
LiH & Li & $-1.3219$ & $3.0763$ & $0.0338$ \\
 & H & $-0.6203$ & $2.1491$ & $1.0550$ \\
\hline
H$_2$O & O & $-8.6489$ & $5.1126$ & $-0.0409$ \\
 & H & $-7.6855$ & $5.4719$ & $-0.0364$ \\
 & H & $-9.0487$ & $5.8301$ & $0.5778$ \\
\hline
BeH$_2$ & Be & $-0.9837$ & $3.4084$ & $0.1009$ \\
 & H & $-1.6817$ & $2.9596$ & $-0.9204$ \\
 & H & $-0.2858$ & $3.8573$ & $1.1222$ \\
\hline
NH$_3$ & N & $-9.0469$ & $4.1587$ & $0.0508$ \\
 & H & $-8.5555$ & $3.4144$ & $0.6423$ \\
 & H & $-8.3995$ & $4.1782$ & $-0.8012$ \\
 & H & $-9.8569$ & $3.5804$ & $-0.3427$ \\
\hline
CH$_4$ & C & $-8.4804$ & $4.2914$ & $0.0000$ \\
 & H & $-9.3089$ & $3.5933$ & $-0.2316$ \\
 & H & $-8.3381$ & $4.3446$ & $1.0974$ \\
 & H & $-7.5481$ & $3.9287$ & $-0.4761$ \\
 & H & $-8.7263$ & $5.2988$ & $-0.3898$ \\
\hline
O$_2$ & O & $-4.9031$ & $4.1890$ & $0.0000$ \\
 & O & $-3.5919$ & $4.1890$ & $0.0000$ \\
\hline
N$_2$ & N & $-6.3552$ & $5.1689$ & $0.0000$ \\
 & N & $-5.1664$ & $5.1830$ & $0.0000$ \\
\hline\hline
\end{tabular}
\end{table}

\subsection{Dynamical Decoupling}
\label{subsec:dd}

Dynamical Decoupling (DD) is an open-loop error suppression technique
with origins in nuclear magnetic resonance (NMR) spin-echo
experiments~\cite{hahn1950spin, carr1954effects}. The quantum computing
adaptation, introduced by Viola and Lloyd~\cite{viola1998dynamical} and
further developed by Viola, Knill, and
Lloyd~\cite{viola1999dynamical} and
Zanardi~\cite{zanardi1999symmetrizing}, suppresses decoherence by
applying sequences of rapid, periodic unitary pulses to idle qubits
during circuit execution. The theoretical mechanism is group-theoretic:
for a system coupled to an environment via
$H_{SE} = \sum_\alpha S_\alpha \otimes B_\alpha$, a DD sequence drawn
from a decoupling group $\mathcal{G}$ causes the time-averaged
system-environment coupling to vanish to leading order in the Magnus
expansion~\cite{viola1999dynamical, zanardi1999symmetrizing}. For a DD
sequence of $N$ pulses applied over a total time $T$ to a qubit subject
to dephasing noise with correlation time $\tau_c$, the residual dephasing
error is suppressed from $\mathcal{O}(T/\tau_c)$ (without DD) to
$\mathcal{O}((T/N\tau_c)^2)$ with DD~\cite{viola1999dynamical}.

Commonly used DD sequences include the Hahn echo ($X$-$X$) and CPMG
($X$-$X$ with optimized timing)~\cite{carr1954effects}. In the present
work, we employ the XpXm DD sequence as implemented in the IBM Quantum
\texttt{Qiskit} framework~\cite{qiskit2023}, which applies alternating
$X^+$ and $X^-$ pulses to idle qubits during circuit execution via the
\texttt{Qiskit} pulse scheduler, inserting $\pi$-pulses into the idle
windows of a dynamically scheduled
circuit~\cite{pokharel2018demonstration, endo2021hybrid}. Experimental
demonstrations on superconducting qubits have confirmed coherence
improvements with DD under appropriate
conditions~\cite{pokharel2018demonstration}; however, the benefit is
contingent on the presence of sufficient idle time. As circuit depth
increases and idle windows shrink relative to active gate time, the
marginal benefit of DD diminishes, and in the deep-circuit regime, DD
may degrade performance due to pulse calibration
errors~\cite{endo2021hybrid}. This depth-dependent behavior motivates
the noise analysis presented in Section~\ref{sec:noise_analysis}, where
the interplay between VQC circuit depth and XpXm DD effectiveness is
characterized in two qualitatively distinct depth regimes.

% =====================================================================
\section{Sidelobe Structure Analysis: Hardware VQC vs.\ Analytical
Dirichlet Reference}
\label{sec:sidelobes}
\label{subsec:results_sidelobes}
% =====================================================================

\begin{figure}[htbp]
    \centering
    \includegraphics[width=\linewidth]{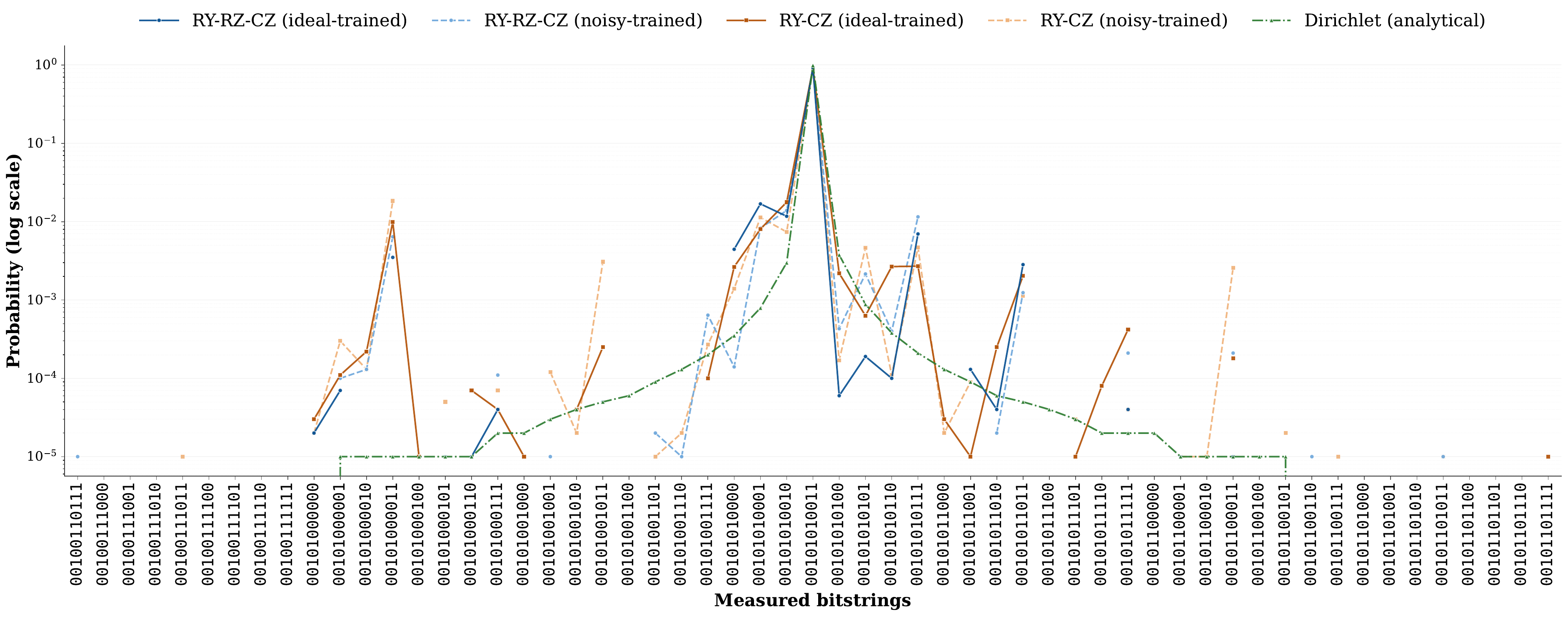}
    \caption{Measured outcome probabilities $P(b)$ (log scale) for the 
    four real-hardware VQC configurations ($R_Y$-$R_Z$-$CZ$ and 
    $R_Y$-$CZ$ ans\"atze, ideal- and noisy-trained, linear entangler, 
    $p=1$) against the analytical Dirichlet reference, plotted against 
    bit-value offset $\Delta(b)$ from the dominant peak 
    \texttt{00101010011} over the window $|\Delta(b)| \le 28$. The 
    Dirichlet distribution decays smoothly with $|\Delta(b)|$, while 
    the hardware distributions exhibit an additional, non-smooth 
    sidelobe at $\Delta(b) = -16$ that is present in all four hardware 
    configurations but negligible in the Dirichlet reference.}
    \label{fig:qpe_distribution_lineplot}
\end{figure}

To characterize how hardware noise distorts the QPE outcome 
distribution near its peak, we compare the four hardware-executed VQC 
configurations against the analytical Dirichlet reference over a 
window of bitstrings centered on the dominant outcome 
$b^{*} = \texttt{00101010011}$. Measurement counts $n(b)$ are converted 
to empirical probabilities $P(b) = n(b) / N$, where $N$ is the total 
number of recorded counts for that configuration ($N = 100{,}000$ for 
all four hardware-executed configurations and $N = 99{,}948$ for the 
analytical Dirichlet reference, whose counts are the truncated expected 
counts described in Section~\ref{subsec:cost}), and each outcome is assigned a 
\emph{bit-value offset} $\Delta(b) = \text{int}(b) - \text{int}(b^{*})$, 
the signed difference between $b$ and $b^{*}$ read as binary integers. 
The plotted window retains $|\Delta(b)| \le 28$ ($57$ outcomes), chosen 
to keep the peak region legible while excluding the sparse tail of the 
full $2^{11}$-outcome space (Figure~\ref{fig:qpe_distribution_lineplot}). 
Within this window we separate two sidelobe types: \emph{near-neighbor} 
outcomes ($0 < |\Delta(b)| \le 5$), where the Dirichlet kernel predicts a 
smooth, monotonically decaying envelope; and \emph{single-bit-flip} 
outcomes at $|\Delta(b)| \in \{8,16\}$, offsets of the form $2^{k}$ that 
include the outcomes reachable from $b^{*}$ by flipping a single bit 
($\Delta(b) = +8$ and $\Delta(b) = -16$), 
which receive no special weight in the Dirichlet target 
and, where they arise from a single bit flip, point to a specific 
physical qubit rather than a nearby phase value.

Table~\ref{tab:sidelobe_summary} summarizes the aggregate sidelobe 
probability, $\sum_{b} P(b)$, for each category. The Dirichlet reference 
shows the smooth decay expected of the Dirichlet kernel: near-neighbor 
outcomes contribute only $\approx 1.0\%$ of the probability mass, and 
single-bit-flip outcomes are essentially unpopulated 
($\approx 0.01\%$). The hardware VQC distributions reproduce the same 
qualitative near-neighbor decay, but with $3$--$4\times$ larger 
aggregate weight ($3.0\%$--$4.0\%$), consistent with broadening of the 
analytical envelope by gate and readout errors. Superimposed on this 
broadened envelope, all four hardware configurations additionally show 
a non-negligible single-bit-flip sidelobe mass ($0.6\%$--$2.5\%$), 
dominated in every case by an isolated peak at $\Delta(b)=-16$ 
($0.35\%$--$1.84\%$) that has no counterpart of comparable size in the 
Dirichlet reference; for the noisy-trained $R_Y$-$CZ$ configuration, 
this single peak exceeds every near-neighbor sidelobe except 
$\Delta(b)=-2$, making it the second most prominent feature in the 
windowed distribution after the dominant peak itself.

\begin{table}[h]
\centering
\caption{Aggregate sidelobe probability (\% of total counts) within 
the $|\Delta(b)| \le 28$ window around \texttt{00101010011}. 
Near-neighbor: $0<|\Delta(b)|\le5$. Single-bit-flip: $|\Delta(b)|\in\{8,16\}$.}
\label{tab:sidelobe_summary}
\renewcommand{\arraystretch}{1.3}
\begin{tabular}{lccc}
\hline\hline
\textbf{Configuration} & \textbf{Dominant} & 
\textbf{Near-neighbor} & \textbf{Single-bit-flip} \\
 & \textbf{peak (\%)} & \textbf{sidelobes (\%)} & \textbf{sidelobes (\%)} \\
\hline
Dirichlet (analytical)          & 98.94 & 0.98 & $\approx$0.01 \\
$R_Y$-$R_Z$-$CZ$, ideal-trained & 92.11 & 4.04 & 0.63 \\
$R_Y$-$R_Z$-$CZ$, noisy-trained & 91.10 & 3.75 & 0.79 \\
$R_Y$-$CZ$, ideal-trained       & 91.50 & 3.68 & 1.24 \\
$R_Y$-$CZ$, noisy-trained       & 90.27 & 3.01 & 2.52 \\
\hline\hline
\end{tabular}
\end{table}

Comparing training environments, noisy-simulator-trained circuits 
consistently show a larger single-bit-flip sidelobe than their 
ideal-trained counterparts for the same ansatz ($R_Y$-$CZ$: $2.52\%$ 
vs.\ $1.24\%$; $R_Y$-$R_Z$-$CZ$: $0.79\%$ vs.\ $0.63\%$), indicating 
that noisy-simulator training does not suppress this qubit-level 
artifact better than ideal-simulator training - if anything, the 
opposite trend holds. Overall, the residual mismatch between the 
hardware VQC outcome distributions and the analytical Dirichlet target 
near the dominant peak arises less from the broadened-but-smooth 
near-neighbor sidelobes, which the VQC reproduces reasonably well in 
shape, and more from a small number of isolated, hardware-specific 
single-bit-flip sidelobes concentrated at a few fixed qubit positions, 
most notably the qubit responsible for $\Delta(b) = -16$.
% =====================================================================
\section{Noise Analysis of VQC Circuit Depth and Dynamical Decoupling}
\label{sec:noise_analysis}
% =====================================================================

As a supplementary investigation beyond the main results of this work,
we present a dedicated noise analysis examining how the interplay between
VQC circuit depth and Dynamical Decoupling (DD) affects the quality of
hardware-executed quantum circuits. This analysis uses the $R_Y$-$CZ$
ansatz with the linear entangler, trained on the ideal simulator, and
executed on IBM Quantum hardware with $N_\text{shots} = 100{,}000$
measurement shots. Motivated by the finding of the
layer-selection study (Section~3.3 of the main paper) that DD consistently
underperforms relative to the non-DD baseline for the $R_Y$-$R_Z$-$CZ$
ansatz across layers $p \in \{1,\ldots,5\}$, we extend this
investigation to a broader range of circuit depths, specifically
$p \in \{8, 64\}$, using the $R_Y$-$CZ$ ansatz, to determine whether the
DD underperformance persists as circuit depth grows, and to identify the
depth range beyond which neither the VQC nor DD can maintain useful
distributional fidelity on current NISQ hardware.

\subsection{Motivation and Experimental Design}
\label{subsec:noise_motivation}

The effectiveness of Dynamical Decoupling as a decoherence-suppression
strategy depends critically on two competing factors: the availability
of idle time windows into which DD pulses can be beneficially inserted,
and the overhead introduced by the DD pulses
themselves~\cite{viola1998dynamical, viola1999dynamical,
pokharel2018demonstration}. For a VQC circuit with $p$ layers of the
linear entangler on $n_a = 11$ qubits, the total circuit time
$T_\text{total}$ is partitioned between active gate time $T_\text{gate}$
and idle time $T_\text{idle} = T_\text{total} - T_\text{gate}$. As $p$
increases, $T_\text{gate}$ grows while the idle-window fraction generally
decreases relative to the total circuit time~\cite{endo2021hybrid},
reducing the opportunity for effective DD insertion.

Two qualitatively distinct depth regimes are investigated:
\begin{itemize}
    \item \textbf{Moderate-depth regime ($p = 8$)}: An intermediate
    depth, substantially deeper than the optimal single-layer case,
    providing a test of whether DD becomes beneficial as the circuit
    lengthens and more idle windows become available.
    \item \textbf{Deep regime ($p = 64$)}: An extreme depth far
    exceeding the coherence length of current IBM Quantum
    superconducting devices, representing the decoherence-dominated
    limit in which the information encoded in the circuit parameters is
    expected to be lost.
\end{itemize}

For each depth, the $R_Y$-$CZ$ ansatz is executed on real IBM Quantum
hardware both with and without the XpXm DD sequence, using
ideal-simulator-trained parameters in both cases. The dominant bitstring
and its count are examined at each depth and DD condition to
characterize the noise behavior.

\subsection{Dominant Bitstring Fidelity Across Depths}
\label{subsec:noise_dominant}

\begin{figure}[htbp]
    \centering
    \includegraphics[width=0.85\linewidth]{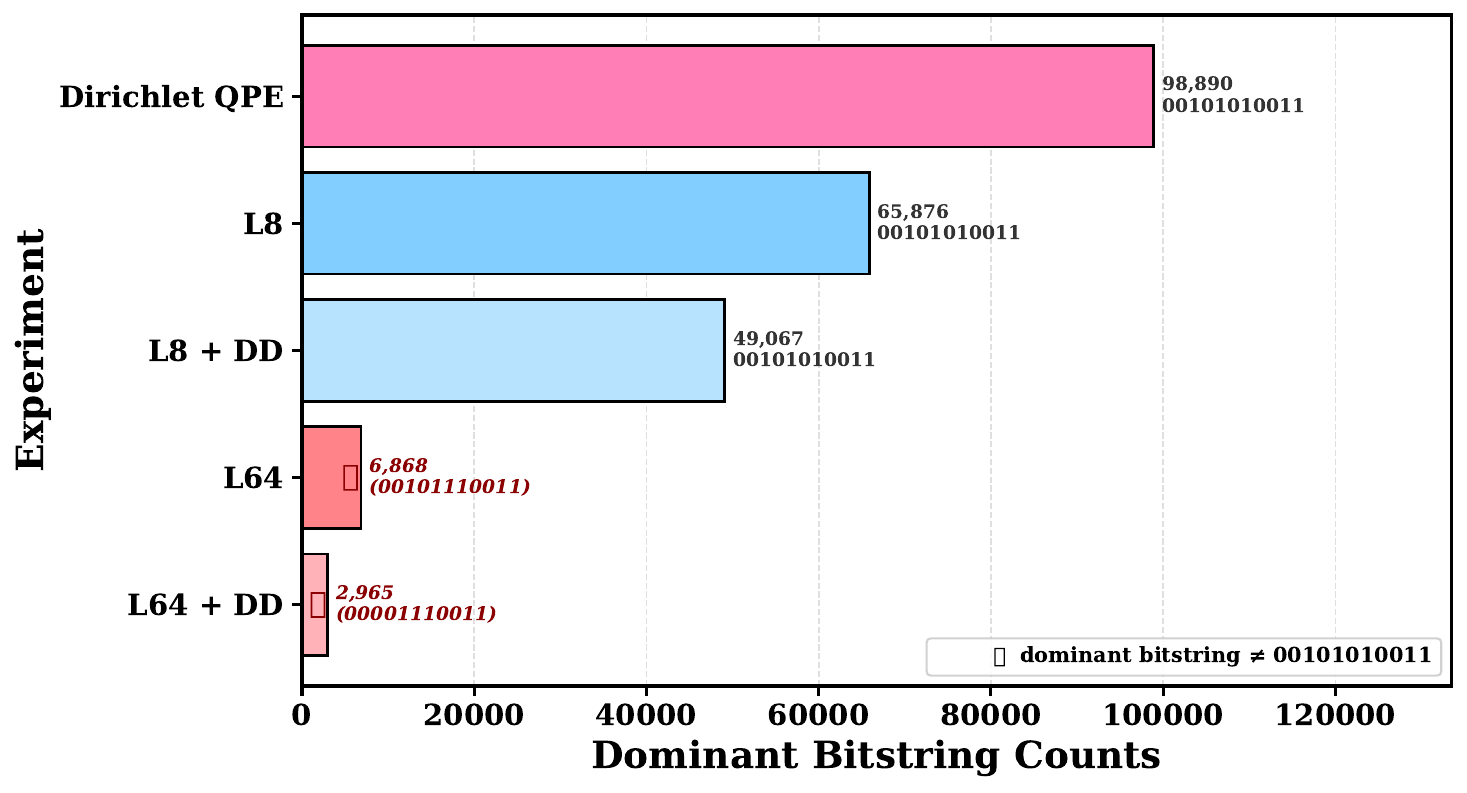}
    \caption{Dominant bitstring counts on real IBM Quantum hardware for
    the $R_Y$-$CZ$ ansatz with linear entangler at layers $p \in \{8,
    64\}$, with and without XpXm DD, compared against the analytical
    Dirichlet reference ($N_\text{shots} = 100{,}000$). At $p = 8$, the
    target bitstring \texttt{00101010011} remains dominant in both the
    DD and non-DD cases. At $p = 64$, the dominant bitstring is no
    longer the target: it shifts to \texttt{00101110011} (non-DD,
    $6{,}868$ counts) and to \texttt{00001110011} (with DD, $2{,}965$
    counts), each flagged in the figure as differing from the target
    bitstring \texttt{00101010011}.}
    \label{fig:noise_dominant_bitstring}
\end{figure}

Figure~\ref{fig:noise_dominant_bitstring} shows the dominant bitstring
and its count for the $R_Y$-$CZ$ ansatz executed on real hardware at
layers $p \in \{8, 64\}$, with and without DD, compared against the
analytical Dirichlet reference count of $98{,}890$ for the target
bitstring \texttt{00101010011}.

At $p = 8$, the dominant bitstring remains the correct target
\texttt{00101010011} in both the DD and non-DD cases, and its count
stays substantial relative to the reference, demonstrating that the
$R_Y$-$CZ$ ansatz with linear entangler retains useful fidelity at this
moderate depth. At $p = 64$, the dominant bitstring is no longer the
target in either case: the non-DD circuit's most frequent outcome is
\texttt{00101110011}, and the DD circuit's most frequent outcome is the
distinct bitstring \texttt{00001110011}. This confirms that, at extreme
depth, the circuit has lost the correct QPE phase information, with the
identity of the (incorrect) dominant outcome itself becoming sensitive
to the presence of DD.

\subsection{Full Distribution Analysis: With and Without Dynamical
Decoupling}
\label{subsec:noise_spread}

To understand how noise and DD interact beyond the dominant bitstring,
we examine the reported counts at $p = 8$ and $p = 64$, comparing
circuits executed with and without the XpXm DD sequence.

\subsubsection*{Layer $p = 8$: DD degrades performance}

At $p = 8$, the non-DD circuit concentrates $65{,}876$ counts on the
target bitstring \texttt{00101010011}, which remains the dominant
outcome, while the DD-augmented circuit reduces this to $49{,}067$
counts, a reduction of approximately $25.5\%$, though
\texttt{00101010011} remains dominant in this case as well. This
reduction is consistent with coherent errors introduced by DD
pulse-calibration imperfections~\cite{pokharel2018demonstration,
endo2021hybrid}, which redistribute probability mass away from the
correct bitstring toward neighboring incorrect ones without displacing
it as the dominant outcome at this depth.

This is a notable finding: even at $p = 8$ layers, substantially deeper
than the optimal single-layer case, the non-DD $R_Y$-$CZ$ circuit still
correctly identifies the target QPE bitstring as dominant, with a
majority-scale count. The linear entangler's sparse two-qubit gate
structure limits cumulative noise accumulation even at this depth,
indicating that the $R_Y$-$CZ$ ansatz with linear entangler is more
robust to hardware noise at moderate depth than a naive circuit-depth
argument would suggest. DD, however, degrades the count on the correct
bitstring at this depth as well, consistent with the DD underperformance
observed for the $R_Y$-$R_Z$-$CZ$ ansatz at $p \in \{1,\ldots,5\}$ in
Section~3.3 of the main paper: taken together, the two
experiments indicate that this underperformance is not confined to a
single ansatz or to shallow circuits, but recurs across different VQC
architectures and into the moderate-depth regime.

\subsubsection*{Layer $p = 64$: both circuits fail}

At $p = 64$, the correct target bitstring \texttt{00101010011} is no
longer the dominant outcome in either case. The non-DD circuit's
dominant bitstring is \texttt{00101110011}, with $6{,}868$ counts, only
a small fraction of the $98{,}890$-count reference; the DD circuit's
dominant bitstring is the distinct \texttt{00001110011}, with $2{,}965$
counts. Both dominant counts are far below the reference, and the
identity of the (incorrect) dominant bitstring differs between the DD
and non-DD cases, indicating that the residual structure of the
measured distribution at this depth is no longer governed by the
learned QPE phase but by depth-dependent noise processes. This is
consistent with the dominant noise mechanism at this depth being
decoherence-driven depolarization rather than the coherent dephasing
that DD is designed to suppress~\cite{viola1998dynamical,
viola1999dynamical, preskill2018quantum}.

This is consistent with the theoretical expectation that, beyond the
circuit-depth threshold set by the hardware coherence time $T_2$,
quantum information is irreversibly lost to the
environment~\cite{preskill2018quantum, leymann2020bitter,
bharti2022noisy}. At $p = 64$ with the linear entangler on $n_a = 11$
qubits, the total circuit depth far exceeds the coherence length of
current IBM Quantum devices, placing this configuration firmly in the
incoherent, noise-dominated regime, in which DD cannot recover useful
distributional fidelity.

\subsection{Depth Regime Characterization}
\label{subsec:noise_threshold}

The two-depth noise analysis reveals a clear progression through
qualitatively distinct operating regimes, summarized in
Table~\ref{tab:noise_summary}, which reports the dominant bitstring and
its count as shown in Figure~\ref{fig:noise_dominant_bitstring}.

\begin{table}[h]
\centering
\caption{Summary of noise analysis results for the $R_Y$-$CZ$ ansatz
with linear entangler (ideal-trained parameters) on real IBM Quantum
hardware, showing the dominant bitstring and its count, with and
without XpXm Dynamical Decoupling, at layers $p \in \{8, 64\}$, as
shown in Figure~\ref{fig:noise_dominant_bitstring}.
$N_\text{shots} = 100{,}000$. Dirichlet reference: target bitstring
\texttt{00101010011}, count $98{,}890$.}
\label{tab:noise_summary}
\renewcommand{\arraystretch}{1.3}
\begin{tabular}{cccccp{3.4cm}}
\hline\hline
\textbf{Layers} $p$ &
\textbf{DD} &
\textbf{Dom.\ Bitstring} &
\textbf{Count} &
\textbf{\% of Ref.} &
\textbf{Regime} \\
\hline
8  & No  & \texttt{00101010011} & 65{,}876 & 66.6\%
& Moderate: correct peak, no DD preferred \\
8  & Yes & \texttt{00101010011} & 49{,}067 & 49.6\%
& Moderate: DD reduces correct-peak weight \\
64 & No  & \texttt{00101110011} & 6{,}868  & 6.9\%
& Deep: wrong peak dominant \\
64 & Yes & \texttt{00001110011} & 2{,}965  & 3.0\%
& Deep: wrong peak dominant, DD provides no benefit \\
\hline\hline
\end{tabular}
\end{table}

\textbf{Moderate-depth regime ($p = 8$):} The non-DD circuit correctly
identifies \texttt{00101010011} as the dominant bitstring, with
$65{,}876$ counts ($66.6\%$ of the reference). This demonstrates the
noise resilience of the sparse linear-entangler $R_Y$-$CZ$ architecture
even at this depth. DD, however, significantly degrades performance,
reducing the count on the (still dominant) correct bitstring by
$\sim 25.5\%$, from $65{,}876$ to $49{,}067$, consistent with coherent
errors introduced by imperfect pulse calibration. This is a practically
important result: the $R_Y$-$CZ$ ansatz can operate usefully at $p = 8$
on current hardware \textit{provided that DD is not applied}. This
underperformance of DD echoes the analogous finding for the
$R_Y$-$R_Z$-$CZ$ ansatz at $p \in \{1,\ldots,5\}$
(Section~3.3 of the main paper), extending the
observation to a second ansatz and a substantially greater depth.

\textbf{Deep regime ($p = 64$):} Both the non-DD and DD-augmented
circuits fail to recover the correct bitstring as dominant. The non-DD
circuit's dominant outcome, \texttt{00101110011}, receives only
$6{,}868$ counts, and the DD circuit's dominant outcome, the distinct
bitstring \texttt{00001110011}, receives only $2{,}965$ counts, both
small fractions of the $98{,}890$-count reference. That the dominant
outcome itself differs between the DD and non-DD circuits, and that
neither recovers the target bitstring, indicates that measurement
statistics at this depth are governed by decoherence-driven noise
rather than the learned QPE phase structure. DD confers no benefit in
this regime.

\subsection{Implications and Design Principles}
\label{subsec:noise_implications}

The noise analysis suggests three practical design principles for
VQC-based QPE learning on current superconducting NISQ hardware.

\textbf{Circuit architecture over error mitigation.} At $p = 8$, the
non-DD $R_Y$-$CZ$ circuit with linear entangler retains substantially
more weight on the correct bitstring than its DD-augmented counterpart.
This indicates that choosing a sparse, hardware-efficient ansatz
architecture is a more effective noise-mitigation strategy than
applying DD as a post-hoc pulse-level intervention. For densely
scheduled VQC circuits, where idle windows are limited, DD introduces
more error than it suppresses.

\textbf{Practical depth limit for current hardware.} On current IBM
Quantum hardware, the $R_Y$-$CZ$ ansatz with linear entangler remains
useful at least up to $p = 8$: within this range, the non-DD circuit
correctly identifies the target bitstring as dominant with a
majority-scale count. By $p = 64$, the dominant bitstring is no longer
the target in either the DD or non-DD case, indicating that the circuit
has entered a decoherence-dominated regime. The precise depth at which
this transition occurs lies between $p = 8$ and $p = 64$ and was not
resolved in this study. This depth limit is hardware-specific and is
expected to improve as device coherence times and gate fidelities
advance~\cite{preskill2018quantum, bharti2022noisy}.

\textbf{DD is not a universal remedy.} DD degrades performance
consistently across every linear-entangler configuration in which it was tested in this
work: for the $R_Y$-$R_Z$-$CZ$ ansatz at $p \in \{1,\ldots,5\}$
(Section~3.3 of the main paper), and for the
$R_Y$-$CZ$ ansatz at $p = 8$ in the present analysis, DD reduces the
count on the correct bitstring relative to the non-DD baseline in every
case where that bitstring remains dominant. At $p = 64$, neither circuit
recovers the correct bitstring as dominant regardless of the pulse
sequence, so DD cannot be said to actively degrade an otherwise-correct
result at this depth so much as to simply fail to help. Taken together,
these results across two ansatz architectures and a depth range spanning
$p = 1$ to $p = 64$ indicate that DD should not be applied as a default
error-mitigation strategy for variational circuits of this kind. Its
benefit is contingent on the presence of substantial and predictable
idle windows, such as those found in quantum error correction circuits
or deliberately padded schedules~\cite{pokharel2018demonstration,
endo2021hybrid}. For the compact linear-entangler VQC circuits studied
here, the idle-window fraction is insufficient to make DD beneficial,
and the additional pulse overhead from DD becomes a source of error
whenever it is applied.

\end{document}